\documentclass[
superscriptaddress,
unsortedaddress,
frontmatterverbose,
onecolumn,
nobibnotes,
amsmath,
amssymb,
aps,
prd,
floatfix,
]{revtex4-2}
\usepackage{amsthm}
\usepackage{dsfont}
\usepackage{color}
\usepackage{graphicx}
\usepackage{dcolumn}
\usepackage{bm}
\usepackage{hyperref}
\usepackage{orcidlink}
\usepackage{subcaption}
\usepackage{booktabs}
\usepackage{float}
\usepackage{tikz}
\usepackage{placeins}
\usepackage{bookmark}
\usepackage{siunitx}
\usepackage{multirow}
\newtheorem{theorem}{Theorem}
\usetikzlibrary{arrows,matrix,positioning}
\begin{document}

\title{High-Order Summation-By-Parts Schemes for First-Order Hyperbolic Systems in Curvilinear Coordinates with Singularities}

\author{Stamatis Vretinaris \orcidlink{0000-0001-7575-813X}}
\email[Corresponding Author: ]{stamatis.vretinaris@aei.mpg.de, stamatis.vretinaris@ru.nl}
\affiliation{Institute for Mathematics, Astrophysics and Particle Physics,
    Radboud University, Heyendaalseweg 135, 6525 AJ Nijmegen, The Netherlands}
\affiliation{Albert-Einstein-Institut, Max-Planck-Institut für Gravitationsphysik, Callinstraße 38, 30167 Hannover, Germany}
\affiliation{Leibniz Universität Hannover, 30167 Hannover, Germany}

\author{Erik Schnetter
    \orcidlink{0000-0002-4518-9017}}
\affiliation{Perimeter Institute for Theoretical Physics, Waterloo, ON N2L 2Y5, Canada}
\affiliation{Department of Physics \& Astronomy, University of Waterloo, Waterloo, ON N2L 3G1, Canada}
\affiliation{Center for Computation \& Technology, Louisiana State University, Baton Rouge, Louisiana 70803, USA}

\date{May 31, 2026}

\begin{abstract}
    Formulating stable numerical methods for hyperbolic systems in curvilinear coordinate with singularities, e.g. spherical coordinates, is complicated by the presence of these singularities. We present a method for constructing high-order accurate, energy-stable finite difference operators satisfying the Summation-by-Parts (SBP) property on spherical domains, extending ideas presented by Gundlach, Martín-García, and Garfinkle \cite{Gundlach2013}. We define discrete gradient and divergence operators that mirror the continuous integration-by-parts principle, even though there is a $1/r^p$ coordinate singularity present at the origin. We explicitly construct such operators up to order six. Our operators place a grid point directly on the origin. We also review how to construct stable SBP operators that straddle the origin. We analyse the accuracy and spectral radii of these operators, and we show example evolutions of the scalar wave equation to demonstrate the advantages of such operators.
\end{abstract}

\maketitle

\section{Introduction}

One often-used method to prove well posedness for a hyperbolic PDE is by constructing a conserved energy (or a non-increasing energy that is bounded from below). The analogue of the well posedness in the continuum is numerical stability for a discretized PDE. Kreiss and Scherer \cite{Kreis1974, Kreiss1977} proposed a powerful way of constructing linearly stable schemes for solving evolution partial differential equations which admit an energy estimate at the continuum level, through the use of difference operators satisfying \emph{summation by parts} (SBP). These operators allow deriving estimates analogous to the continuum ones, up to boundary terms.

While the former guarantees well posedness, the discrete counterparts guarantee numerical stability. Numerical stability prevents unbounded growth, i.e. prevents exponential growth with a resolution-dependent growth rate where higher resolutions have faster growing modes.

Summation by parts is the discrete equivalent of integration by parts. A set of finite differencing operators may or may not satisfy the SBP property; if they do, they allow certain kinds of proofs and estimate to ``go through'' for the discretized equations in the same way as for the continuum equations. For example, if a hyperbolic partial differential equation (PDE) has a conserved energy, then discretizing this system using SBP operators often leads to discrete system where the discrete energy is also conserved. This then holds \emph{independent of the resolution}, i.e. no matter how large the discretization error -- energy conservation is then not a property obtained in the limit of infinite resolution, but it can be tested at any resolution, up to floating-point error.

There is a large body of literature for SBP operators that we briefly review further below. In this paper we will follow ideas first presented by Gundlach, Martín-García, and Garfinkle \cite{Gundlach2013} and construct SBP finite differencing operators (stencils) for curvilinear coordinate systems with singularities, for example for polar or spherical coordinates. We introduce a set of \emph{accuracy conditions} that our operators need to satisfy near the origin, and then explicitly construct such operators for fourth and sixth order accuracy. Our operators retain their full accuracy at the origin; in fact, we are even able to demand a somewhat higher accuracy near the origin to select from a family of operators.

It goes without saying that spherical symmetry, or multipole expansion in spherical coordinates, or in general curvilinear coordinate systems with singularities are immensely useful in describing many physical systems. One famous example in general relativity is Choptuik's study of gravtitational critical collapse \cite{Choptuik1993,Gundlach2007,Gundlach2025}. Many systems in general relativity exhibit symmetries that can be well captured by a curvilinear coordinate system, such as supernova core collapse simulations \cite{Muller2020}, black hole mimickers like boson stars \cite{Liebling2017}, black strings \cite{Lehner2010}, etc.

The method we describe in this paper has the potential to improve accuracy and guarantee discrete stability in all cases where high-order finite differences are applicable.

Issues related to coordinate singularities have been studied outside numerical relativity. Prochnow et al. proposed a high-order SBP treatment for the polar singularity in axisymmetric wave propagation using staggered SBP operators; their method avoids evolving the unknown at r=0 and uses weighted SBP norms so that the origin contributes no spurious boundary work \cite{Prochnow2017}. O’Reilly and Petersson developed energy-conservative SBP discretizations for acoustic wave propagation in covariant form on staggered curvilinear grids \cite{OReilly2020}. Mimetic finite-difference methods provide another staggered-grid framework with discrete divergence/gradient operators and compatible quadratures \cite{Srinivasan2023}. Other approaches include geometric transformations or pole conditions that avoid direct evaluation at the polar point \cite{Mohseni2000}, centerline equations obtained with l’Hospital’s rule \cite{Griffin1979}, series-expansion treatments of cylindrical-axis singularities \cite{Constantinescu2002}, and specialized finite-element basis modifications at polar grid centers \cite{Bhole2022}.

Since the singularity at the origin is not physical, but a mere pathology of the coordinate system, it is possible to construct appropriate SBP operators that treat this pathology. This was first shown in \cite{Gundlach2013}, where the authors constructed stable second and fourth order SBP operators for spherical decompositions of the wave equation.

In this paper we extend the work of \cite{Gundlach2013} on spherical/curvilinear SBP operators and develop a general method to construct SBP operators of arbitrary order that do not lose accuracy at coordinate singularities. We formulate our operator construction by demanding \emph{accuracy conditions}, which are linear conditions on SBP the operators, generalizing the recurrence relations introduced in \cite{Gundlach2013}.
We rely on existing Cartesian SBP operators, and then transform these operators to be applicable in curvilinear coordinate systems, finding appropriate origin closures for the norm operators such that the resulting operators are stable and accurate.

We give a general overview of the summation-by-parts idea and its relevance to discrete stability in section \ref{sec:theory}. We also review some basic relevant properties of curvilinear coordinates and tensor multipole moments there. In section \ref{sec:sbp_framework_curvilinear} we describe our construction and show two examples for fourth- and sixth-order accurate SBP operators for spherical symmetry. Finally, we show example evolutions of the scalar wave equation with these operators in section \ref{sec:numerical}, demonstrating the expected convergence and accuracy, in particular at and near the coordinate singularity at the origin.

\section{Summation By Parts}
\label{sec:theory}

\subsection{Integration By Parts}

Give a scalar function $\Pi$ and a vector-valued function $\Psi$ on some domain $\Omega$ in a suitable function space, the well-known \emph{integration by parts} theorem states that
\begin{eqnarray}
    \int_\Omega (\mathrm{grad}\, \Pi) \cdot \Psi\, dV + \int_\Omega \Pi\, (\mathrm{div}\, \Psi)\, dV &=& \oint_{\partial\Omega} \Pi\, \Psi \cdot ds
\end{eqnarray}

Integration by parts often plays a useful role in proving the well-posedness of wave-type hyperbolic partial differential equations (PDEs). For example, the wave equation
\begin{eqnarray}
    \label{eqn:wave}
    \partial_t \Pi &=& \mathrm{grad}\, \Psi \\
    \partial_t \Psi &=& \mathrm{div}\, \Pi
\end{eqnarray}
has the conserved energy
\begin{eqnarray}
    E &=& \frac{1}{2} \int_\Omega \Pi^2 + \Psi^2\, dV
\end{eqnarray}

A sufficient condition for this PDE to be well-posed is the existence of a conserved (or non-increasing) energy \cite{Friedrichs1954, Friedrichs1928, Friedrichs1958} (which is bounded from below). We show that this is the case here:
\begin{eqnarray}
    \partial_t E &=& \int_\Omega \Pi\, (\partial_t \Pi) + \Psi \cdot (\partial_t \Psi)\, dV \\
    &=& \int_\Omega \Pi\, (\mathrm{div}\, \Psi) + \Psi \cdot (\mathrm{grad}\, \Pi)\, dV
\end{eqnarray}
and use integration by parts to find
\begin{eqnarray}
    E &=& \oint_{\partial\Omega} \Pi\, \Psi \cdot ds\, .
\end{eqnarray}
Given suitable boundary conditions, this energy is conserved (in particular for either $\Pi=0$ or $\Psi=0$) or non-increasing.

We can also write the integration-by-parts theorem in operator notation as
\begin{eqnarray}
    \langle \mathrm{grad}\, \Pi| H_V | \Psi \rangle + \langle \Pi | H_S | \mathrm{div}\, \Psi \rangle &=& \langle \Pi | B | \Psi \rangle
\end{eqnarray}
where we allow for a non-trivial volume element (integration weight, norm) $H_S$ and incorporate the dot product into $H_V$. $\mathrm{bnd}$ is an operator describing the integral over the boundary. This can also be written concisely as operator equation
\begin{eqnarray}
    \mathrm{grad}^T H_V + H_S \mathrm{div} &=& \mathrm{bnd}\, .
\end{eqnarray}
It is now evident that integration by parts is a condition on the derivative operators (gradient and divergence), norms (for scalar and vector-valued functions), and boundary operator.

We remark that the gradient is always given by a partial derivative, whereas the divergence as understood in this paper is the covariant divergence and depends on the metric of the manifold.

\subsection{Summation By Parts}

\emph{Summation by parts} is the discrete equivalent of integration by parts, i.e. it is applicable to discretized functions, and is similarly useful there. A sufficient condition for stability of a spatial discretization of a PDE is the existence of a discrete energy estimate, namely a conserved or non-increasing discrete energy. This approach is commonly referred to as the energy method \cite{Gustafsson2013, Strand1994}.

We discretize (\ref{eqn:wave}) in space but leave time continuous. We use an asterisk $\Box^*$ to indicate discrete quantities and discrete operators.

This system is then
\begin{eqnarray}
    \partial_t \Pi^* &=& \mathrm{grad}^*\, \Psi^* \\
    \partial_t \Psi^* &=& \mathrm{div}^*\, \Pi^*
\end{eqnarray}
with the discrete energy
\begin{eqnarray}
    E^* &=& \frac{1}{2} \left( \Pi^*\,H_S^*\,\Pi^* + \Psi^*\,H_V^*\,\Psi^* \right)\, .
\end{eqnarray}
The expressions for the energy corresponds to the integral above, written in operator notation. The operators $H_S^*$ and $H_V^*$ are discrete quadrature rules \cite{DelReyFernandez2014b}.

If we now demand that our discrete operators satisfy the summation-by-parts property
\begin{eqnarray}
    (\mathrm{grad}^*)^T H_V ^* + H_S^* \mathrm{div}^* &=& \mathrm{bnd}^*
\end{eqnarray}
then this discrete energy is also conserved up to boundary terms:
\begin{eqnarray}
    \partial_t E^* &=& \Pi^* \mathrm{bnd}^* \Psi^*\, .
\end{eqnarray}
The proof is exactly equivalent to the continuum proof above. This is the main feature of summation-by-parts operators: They allow translating a continuum proof to a proof for a discrete system, and this simplifies finding a stable discretization for a well-posed hyperbolic PDE.

\subsection{Spherical Coordinates and Tensor Rank}
\label{subsec:effective_measure}

In spherical coordinates $x^i = (r, \theta^A)$ in $d$ dimensions, the volume element is $dV = r^{d-1} \sqrt{\det \Omega} \, dr \, d^{d-1}\theta$, where $\Omega_{AB}$ is the metric on the unit sphere $S^{d-1}$. For a physical tensor field $\mathbf{\Phi}$, the continuous $L^2$ energy norm is defined by:
\begin{equation}
    \|\mathbf{\Phi}\|^2 = \int_{\Sigma} |\mathbf{\Phi}|^2 \, r^{d-1} \, dr \, d\Omega
\end{equation}

When the angular dependence of the field is projected onto a basis of tensor (hyper)spherical harmonics of degree $l$, the radial profile of a rank-$s$ tensor multipole scales with a leading power $r^k$ as $r \to 0$. (This is necessary so that its Cartesian tensor components remain smooth).

For a generic tensor field, $k$ takes the form
\begin{equation}
    k = l + \delta_s
\end{equation}
where $l$ is the multipole degree, and the integer shift $\delta_s \in \{-s, -s+1, \dots, s\}$ is determined by the tensor rank $s$ and the specific parity of the tensor harmonic component (e.g., electric vs. magnetic parity). For a scalar field ($s=0$), the shift is $\delta_0 = 0$, yielding $k=l$.

To construct a numerical scheme that automatically enforces this required analyticity, we define a regularized tensor variable $\mathbf{\tilde{\Phi}}$ by factoring out the geometric scaling:
\begin{equation}
    \mathbf{\Phi}(r) = r^k \mathbf{\tilde{\Phi}}(r)
\end{equation}

We substitute the regularized variable back into the continuous $L^2$ energy norm and find
\begin{align}
    \|\mathbf{\Phi}\|^2 & = \int (r^k \mathbf{\tilde{\Phi}})^2 \, r^{d-1} \, dr \, d\Omega\nonumber \\
                        & = \int \mathbf{\tilde{\Phi}}^2 \, r^{2k + d - 1} \, dr \, d\Omega
    \label{eq:regularized-energy-norm}
\end{align}
Equation (\ref{eq:regularized-energy-norm}) implies that factoring out the analytic regularity condition maps the dynamics of the smooth variable $\mathbf{\tilde{\Phi}}$ into a new effective function space with the modified volume element $\tilde\gamma$:
\begin{equation}
    \tilde\gamma\, dr = r^{2k + d - 1}\, dr = r^p\, dr
\end{equation}
where we define $p = 2k+d-1$.

In our first-order symmetric hyperbolic system, the evolution is governed by the gradient and divergence operators. We define here a new set of operators that act on the regularized variables. These operators are rescaled versions of their Cartesian counterparts, and are defined as follows:
\begin{eqnarray}
    \tilde{G} &=& \mathrm{grad} \\
    \tilde{S} &=& r^p H_S \\
    \tilde{V} &=& r^p H_V \\
    \tilde{B} &=& r^p \mathrm{bnd}
\end{eqnarray}
We demand that the rescaled operators should satisfy integration by parts
\begin{eqnarray}
    \tilde{S} \tilde{D} + \tilde{G}^T \tilde{V} &=& \tilde{B}\, ,
\end{eqnarray}
and the rescaled divergence operator can thus be defined via
\begin{align}
    \tilde{D} & = \tilde{S}^{-1} \left( \tilde{B} - \tilde{G}^T \tilde{V} \right)\, .
\end{align}
where we used the fact that $\tilde{H_S}$ is a norm and thus invertible.

Expanded and simplified, this leads to
\begin{eqnarray}
    \tilde{G} &=& \partial_r \\
    \tilde{D} &=& \partial_r + \frac{p}{r}\, .
\end{eqnarray}
The term $\frac{p}{r}$ is the expected Christoffel symbol, which is singular at the origin. The main point of designing summation-by-parts operators is to ensure that the formal singularities in the divergence operator are guaranteed to cancel not only in the continuum but also in the discrete, so that a discrete divergence operator can be applied without worrying about the origin. As the equation above suggests, this means that our set of discrete operators will  contain a (covariant) divergence operator, and not separate partial derivatives and Christoffel symbols.

The parameter $p$ combines all of the spatial dimension $d$, the multipole degree $l$, and the tensor rank $s$ into a single integer. For example, applying this to the 3D scalar wave equation ($d=3, \delta_0=0$) leads to the expected $p = 2l + 2$ \cite{Gundlach2013}. In the next section we develop a method to construct SBP operators that handle the singular $p/r$ term in the divergence operator and lead to a stable discretization, applicable to any tensorial hyperbolic system, in any spatial dimension, for any multipole expansion.

\subsection{Order and Degree of Derivative Operators}
\label{sec:order-and-degree}

For classical finite-difference summation-by-parts (FD-SBP) operators, the \emph{degree} of the operator is defined as the maximum degree of the monomials that are differentiated exactly \cite{DelReyFernandez2015}, i.e. without any discretization error. The discrete derivative approximation introduces a leading truncation error at each node that scales with some power of the grid spacing $h$. The \emph{order} of the operator is defined by the minimum exponent of $h$ that characterizes these local truncation errors. For an operator approximating the $m$-th derivative, the relationship between order and degree is
\begin{equation}
    \text{order} = \text{degree} - m + 1.
\end{equation}
Because we are here primarily concerned with first derivatives ($m=1$), the order of accuracy and the degree of exactness are equivalent for classical FD-SBP operators in Cartesian coordinates.

However, this direct equivalence does not persist when transforming to curvilinear coordinates. While the gradient operator retains this property, the exactness degree of the divergence operator changes. The relationship becomes
\begin{equation}
    \text{degree} = \text{order} - p
    \label{eq:sbp-degree}
\end{equation}
where $p$ is the geometric parameter introduced above. Although both the gradient and divergence operators share the same formal order of accuracy, their polynomial degrees of exactness differ. This discrepancy arises from the metric term $r^p$ in the continuous (covariant) divergence operator  (e.g., $p=1$ for cylindrical coordinates and $p=2$ for spherical coordinates):
\begin{equation}
    \mathcal{D} f(r) = \frac{1}{r^p} \frac{\partial}{\partial r} \left( r^p f(r) \right).
\end{equation}
When applying $\mathcal{D}$ to a monomial $f(r) = r^k$, the derivative $\partial_r$ acts on $r^{k+p}$. Consequently, for the divergence operator to be exact for a polynomial of degree $k$, the underlying first-derivative FDSBP operator must be exact for polynomials up to degree $k+p$, thereby lowering the effective exactness degree of the operator by $p$. Assuming an FD stencil of order $q$, the resulting divergence operator remains $q$-th order accurate, while its polynomial exactness is reduced according to Eq. (\ref{eq:sbp-degree}).

\section{Constructing Summation-by-Parts Operators for Curvilinear Coordinates with Singularities}
\label{sec:sbp_framework_curvilinear}

As described in the previous section, proving the energy stability of a hyperbolic PDE relies on the fact that the spatial derivative operators, i.e. the gradient $\tilde{G} = \partial_r$ and the weighted divergence $\tilde{D} = \partial_r + p/r$ satisfy the integration-by-parts property with respect to the volume element $r^p$.

Summation-by-Parts (SBP) operators achieve discrete stability by ensuring that the discrete matrix operators satisfy an algebraic analog.

For simplicity, we now drop the tilde for rescaled operators/quantities and the asterisk for discrete operators/quantities.

\subsection{Discrete Norm Operators}

We use finite differences to construct the discrete operators. We map the radial domain $r \in [0, R]$ onto a grid with $N$ nodes, $r_i = (i-1)\, h$. This places the first grid point $i=1$ exactly at the origin.

For simplicity and elegance we will present our method for non-staggered grids, i.e. for grids which place one grid point exactly at the origin. Our method can also be applied to staggered grids.

The origin ($r=0$) becomes a boundary point in our discrete domain. This is not a physical boundary. The boundary conditions there are given by symmetry conditions. Different variables satisfy different symmetry conditions there:
\begin{itemize}
    \item \textbf{Scalar fields:} Regularized scalars are even functions ($\Pi(-r) = \Pi(r)$), and their value at the origin is non-zero in general ($\Pi(0) \neq 0$). Their gradient is odd and vanishes at the origin.
    \item \textbf{Vector fields:} Regularized vectors are odd functions ($\Psi(-r) = - \Psi(r)$), and they vanish at the origin. Their gradient is even and does in general not vanish at the origin.
\end{itemize}

Above, we already used a two different norm operators (mass matrices) $H_S$ and $H_V$ for the scalar and vector components in our state vector. Here we also define two different discrete SPD norm operators for scalar and vector fields, respectively \cite{Gundlach2013}:
\begin{itemize}
    \item \textbf{The Scalar Norm Operator ($S$):} Defines the discrete inner product for the even parity fields.
    \item \textbf{The Vector Norm Operator ($V$):} Defines the discrete inner product for odd parity fields. Because regularized vectors vanish at the origin, their geometric energy contribution at the origin is zero. This decouples the  weight at the origin $V_{11}$ from the standard geometric quadrature rules. In principle one could remove the value at the origin from the state vector for odd functions. We do not do that because this would increase the complexity of our setup for very little practical gain.
\end{itemize}
In the one-dimensional Cartesian case there is only a single norm $H$ because there is no distinction between scalars and vectors and between gradient and divergence.

Since we include a grid point at $r=0$, it is tempting to choose the value of $S$ at the origin $S_{11} = 0$. This would make $S$ singular; it would violate the positivity condition $S$ needs to satisfy to be a norm. It is therefore necessary to regularize the operators. To resolve this while preserving discrete stability, we construct a modified scalar norm matrix where the weight at the origin is positive, $S_{11} = \mathcal{O}(h^{p+1}) > 0$. This regularized weight guarantees that $S$ is invertible (which will necessary to construct the divergence operator). It also maintains the formal order of accuracy of volume integration.

\subsection{Discrete Derivative Operators}

With these parity-specific norms, the continuous derivatives $\mathrm{grad}$ and $\mathrm{div}$ are approximated by discrete operators $G$ and $D$. The generalized Summation-by-Parts property then demands
\begin{eqnarray}
    S D + G^T V &=& B
    \label{eq:curvilinear-sbp}
\end{eqnarray}
where $B = \text{diag}(0, \dots, 0, R^p)$ is a diagonal matrix that projects onto the outer boundary point.

We choose to construct the discrete divergence operator $D$ from the other operators using the generalized curvilinear SBP relation (\ref{eq:curvilinear-sbp}):
\begin{eqnarray}
    \label{eq:div_from_sbp}
    D &=& S^{-1} (B - G^T V)\, .
\end{eqnarray}
That is, we first choose the gradient $G$, the norms $S$ and $V$, and the boundary operator $B$, and then construct the divergence $D$ from these, ensuring that the SBP property holds. Since $S$ is a norm, it is by definition symmetric and positive and thus invertible.

To construct the discrete derivative operators, we start by defining the $s$-th order accurate cartesian SBP operator $G$. We do so by starting with a given Cartesian SBP operator. Near the origin, where the stencil coefficients would naively reach ``across the origin'' into negative radii, we use the symmetry of scalar functions $\Pi(r)$ to map the operator back into the domain. It is important to notice that we do \emph{not} treat the origin as a boundary. The behavior of $G$ near the origin is defined only in terms of the symmetry of scalar functions. We also note in passing that the gradient is an operator that maps scalar functions to vector-value functions while the divergence maps vector-valued functions back to scalar functions.

We also apply this symmetry condition to the Cartesian norm $H$, which in practice means that the matrix element $H_{11}$ picks up a factor of $1/2$ compared to its Cartesian value. $H_{11}$ is not directly applicable to our domain because it does not take the volume element into account. It is only a starting point for defining the norm operators $S$ and $V$.

Finally we define our boundary operator $B$ to be the Cartesian boundary operator, scaled by $R^p$.

\subsection{Non-Diagonal Norm Operators}

It appears to be impossible to find a consistent set of higher-order operators $G$, $D$, $S$, $V$, and $B$ that satisfy summation-by-parts while keeping the norm operators diagonal, if one also insists that the operators retain their degree of exactness and convergence order near the origin. (We tried and failed -- we have not proved that this is impossible.) Since the origin is ``difficult'' to handle anyway, it seems prudent to us to insist that our scheme should be as accurate as possible there, and we do not want to loosen this requirement. Instead we thus choose to allow the vector norm operator to have a few off-diagonal elements near the origin, leading to \emph{non-diagonal norm operators}.

In practice this means that there are cross-terms when integrating over the domain. For example, a term $V_{2,3}$ means that the norm $\Psi V \Psi$ contains not only terms such as $V_2 \Psi_2^2 + V_3 \Psi_3^2$, but that there is also an additional term $\Psi_2 V_{2,3} \Psi_3$ contributing to the norm.

These off-diagonal elements have very little effect in practice. A typical evolution scheme will only use gradient, divergence, and boundary operators, and these retain the usual structure. The norm operators are mostly relevant if one wants to integrate over the domain, e.g. to calculate the total energy. This is usually not part of the evolution scheme, but is only used as analysis method, and the small additional complication posed by a few non-diagonal terms do not pose a large programming burden.

Since we compute the divergence from the SBP relation, which contains the inverse of the scalar norm, we keep the scalar norm $S$ simple and diagonal.

\begin{eqnarray}
    S &=& h\, \mathrm{diag}(s_1, \ldots ,s_N)
\end{eqnarray}
The values of $s_i$ will be determined below.

We construct a non-diagonal vector norm $\mathrm{V}$ as described in the following subsections. Given that $\mathrm{V}$ is defined for vector-valued functions, which are odd functions and are thus zero at the origin, the elements of the first row and column of $\mathrm{V}$ do not matter since they are always applied to a value of $u_1 = u(0) = 0$. We choose to set  $\mathrm{V}_{1,1} = 1$ for simplicity.

Below, we construct the operators while setting $h=1$ for simplicity. (It is straightforward to rescale them by the respective factors $h^p$.)

Simple and convenient second order operators were constructed in \cite{Gundlach2013}. We do not describe second order operators here and start by constructing fourth-order operators.

\subsubsection{4th order accurate non-diagonal norm operators}

For a fourth order accurate vector norm $\mathrm{V}$ we use the ansatz
\begin{equation}
    \mathrm{V} = \begin{matrix}
        \begin{tikzpicture}
            \matrix [matrix of math nodes,left delimiter=(,right delimiter=)] (m)
            {
                1      & 0       & 0       & 0       & 0       & 0      & \cdots & 0      \\
                0      & v_{2}   & v_{2,3} & 0       & 0       & 0      & \cdots & 0      \\
                0      & v_{2,3} & v_{3}   & v_{3,4} & 0       & 0      & \cdots & 0      \\
                0      & 0       & v_{3,4} & v_{4}   & v_{4,5} & 0      & \cdots & 0      \\
                0      & 0       & 0       & v_{4,5} & v_{5}   & 0      & \cdots & 0      \\
                0      & 0       & 0       & 0       & 0       & s_{6}  & \cdots & 0      \\
                \vdots & \vdots  & \vdots  & \vdots  & \vdots  & \vdots & \ddots & \vdots \\
                0      & 0       & 0       & 0       & 0       & 0      & \cdots & s_{n}  \\
            };
            \draw[rounded corners,ultra thick, draw=black, fill=blue, opacity=0.3] ([xshift=-2pt,yshift=-2pt]m-5-1.south west) rectangle ([xshift=3pt,yshift=2pt]m-1-5.north east);
        \end{tikzpicture}
    \end{matrix}
    \label{eq:vector-norm-4}
\end{equation}

We add a small region with one additional symmetric non-zero band near the origin. Further away from the origin we set $\mathrm{V} = \mathrm{S}$. A similar off-diagonal structure is used in \cite{Gundlach2013} for the 4th order operators presented there.

To determine the values of $s_i$ and $v_{ij}$, we first formally calculate the divergence $\mathrm{D}$ from the SBP relation, and we then impose the following \emph{accuracy constraints}:
\begin{enumerate}
    \item Everywhere, except close to the outer boundary, we demand that the divergence is exact when applied to the linear function $f(r)=r$, i.e. $\mathcal{D}\, r = p+1$. We cannot demand this near the outer boundary because the underlying Cartesian SBP gradient loses accuracy near the outer boundary.
    \item Near the origin we also demand that the divergence is exact for $f(r)=r^3$. We do this to increase the number of conditions to match the number of unknowns (see below).
    \item Finally we demand that the volume of the domain is exact, i.e. that $\int f(r)\, r^p\,dr$ is exact for $f(r)=1$.
\end{enumerate}
We do so because the first and third conditions are special cases, and it seems advantageous to handle them exactly.

We find a sufficient number of off-diagonal non-zero elements by trial and error. The ansatz here leads to a fourth-order operator with good properties. It is certainly possible to allow for more off-diagonal elements, and we do so for defining a sixth-order operator in the next section.

In detail, the accuracy conditions above lead to the equations
\begin{align}
    (\mathrm{D} r)_i              & = p+1  \quad \quad \quad i=1 \ldots N-N_c \\
    (\mathrm{D} r^3)_i            & = (p+3) \, r^2 \quad i=1 \ldots 5         \\
    \int_0^R \mathbf{1}\,r^p\, dr & = \sum_i s_i = \frac{R^{p+1}}{p+1}
\end{align}
where $N_c$ is the width of boundary closure (i.e. the number of ``special'' points near the boundary) of the underlying Cartesian operators.

The operator $\mathrm{D}$ above is defined as $\mathrm{D} = \mathrm{S}^{-1} (\mathrm{B} - \mathrm{G}^T \mathrm{V})$, where $\mathrm{G}$ and $\mathrm{B}$ are known from their respective Cartesian operators. We insert this definition into the conditions above and then multiply both sides by $\mathrm{S}$, leading to
\begin{eqnarray}
    (\mathrm{B}^i{}_j - \mathrm{G}_k{}^i \mathrm{V}^k{}_j)\, r^j &=& (p+1)\, \mathrm{S}^i{}_j\, 1^j \quad\quad i=1 \ldots N-N_c \\
    (\mathrm{B}^i{}_j - \mathrm{G}_k{}^i \mathrm{V}^k{}_j)\, (r^3)^j &=& (p+3)\, \mathrm{S}^i{}_j\, (r^2)^j \quad i=1 \ldots 5 \\
    1_i\, S^i{}_j\, 1^j &=& \frac{R^{p+1}}{p+1}
\end{eqnarray}
where $1^i$, $r^i$, $(r^2)^i$, $(r^3)^i$ are vectors of grid point values for the functions $f(r)=1$, $f(r)=r$, etc.

The only unknowns in these equations are the values $s_i$ and $v_{ij}$. This system of equations is linear and can be solved in a straightforward manner. We solve this system using rational numbers (not floating-point numbers) to ensure we find an exact solution.

One needs to be careful when counting the number of linearly independent equations because $r^i$ and $(r^3)^i$ are zero at the origin, reducing the number of linearly independent equations. For this fourth-order operator we demand third-order exactness at $5$ points near the origin to ensure that the number of linearly independent equations matches the number of unknowns in $\mathrm{S}$ and $\mathrm{V}$. If one wanted to modify our ansatz, then this number $5$ might need to be modified accordingly.

Perhaps surprisingly, the solution for the scalar norm $\mathrm{S}$ is $s_i = r_i^p H_i$. These are exactly the values that one obtains by scaling the Cartesian norm $\mathrm{H}$ by the volume element. The scalar norm has non-trivial entries near the origin (as expected) and near the outer boundary (where the Cartesian operators have a reduced accuracy).

As example, we show the resulting operators for a domain size of $R=30$ and with a grid spacing $h=1$ (for simplicity), in spherical symmetry (i.e. $p=2$), using the Mattsson et. al. operators \cite{Mattsson2004a}: The Cartesian norm is
\begin{eqnarray}
    \mathrm{H} &=& \operatorname{diag}\!\left(\frac{1}{2},1,\dots,1, \frac{49}{48}, \frac{43}{48}, \frac{59}{48}, \frac{17}{48} \right)
\end{eqnarray}
Our scalar norm:
\begin{eqnarray}
    \mathrm{S} &=& \operatorname{diag}\!\left(
    \dfrac{3714185}{6311896},
    \dfrac{876978}{788987},
    \dfrac{2673132}{788987},
    \dfrac{7012892}{788987},
    \dfrac{101138679}{6311896},
    5^2,6^2,\dots,26^2,
    \dfrac{11907}{16},
    \dfrac{2107}{3},
    \dfrac{49619}{48},
    \dfrac{1275}{4} \right)
\end{eqnarray}
Our vector norm:
\begin{eqnarray}
    \mathrm{V} &=& \begin{matrix}
        \begin{tikzpicture}
            \matrix [matrix of math nodes,left delimiter=(,right delimiter=)] (m)
            {
                1 & 0                       & 0                       & 0                           & 0                         \\
                0 & \dfrac{2689591}{788987} & \dfrac{8002}{46411}     & 0                           & 0                         \\
                0 & \dfrac{8002}{46411}     & \dfrac{3706327}{928220} & \dfrac{904416}{3944935}     & 0                         \\
                0 & 0                       & \dfrac{904416}{3944935} & \dfrac{251857497}{27614545} & -\dfrac{1132080}{5522909} \\
                0 & 0                       & 0                       & -\dfrac{1132080}{5522909}   & \dfrac{89215604}{5522909} \\
            };
            \draw[rounded corners,ultra thick, draw=black, fill=blue, opacity=0.3] ([xshift=-2pt,yshift=-10pt]m-5-1.south west) rectangle ([xshift=22pt,yshift=2pt]m-1-5.north east);
        \end{tikzpicture}
    \end{matrix}
    \oplus
    \operatorname{diag}\!\left(
    5^2,6^2,\dots,26^2,
    \dfrac{11907}{16},
    \dfrac{2107}{3},
    \dfrac{49619}{48},
    \dfrac{1275}{4}
    \right)
\end{eqnarray}

\subsubsection{6th order accurate non-diagonal norm operators}

To construct a sixth-order operator, we start with a similar ansatz for the vector norm $\mathrm{V}$, with more non-zero off-diagonal elements:
\begin{equation}
    \mathrm{V} = \begin{pmatrix}
        1      & 0       & 0       & 0       & 0       & 0       & 0       & 0       & 0       & 0         & 0      & \cdots & 0      \\
        0      & v_{2,2} & v_{2,3} & v_{2,4} & 0       & 0       & 0       & 0       & 0       & 0         & 0      & \cdots & 0      \\
        0      & v_{2,3} & v_{3,3} & v_{3,4} & v_{3,5} & 0       & 0       & 0       & 0       & 0         & 0      & \cdots & 0      \\
        0      & v_{2,4} & v_{3,4} & v_{4,4} & v_{4,5} & 0       & 0       & 0       & 0       & 0         & 0      & \cdots & 0      \\
        0      & 0       & v_{3,5} & v_{4,5} & v_{5,5} & v_{5,6} & 0       & 0       & 0       & 0         & 0      & \cdots & 0      \\
        0      & 0       & 0       & 0       & v_{5,6} & v_{6,6} & v_{6,7} & 0       & 0       & 0         & 0      & \cdots & 0      \\
        0      & 0       & 0       & 0       & 0       & v_{6,7} & v_{7,7} & 0       & 0       & 0         & 0      & \cdots & 0      \\
        0      & 0       & 0       & 0       & 0       & 0       & 0       & v_{8,8} & 0       & 0         & 0      & \cdots & 0      \\
        0      & 0       & 0       & 0       & 0       & 0       & 0       & 0       & v_{9,9} & 0         & 0      & \cdots & 0      \\
        0      & 0       & 0       & 0       & 0       & 0       & 0       & 0       & 0       & v_{10,10} & 0      & \cdots & 0      \\
        0      & 0       & 0       & 0       & 0       & 0       & 0       & 0       & 0       & 0         & s_{11} & \cdots & 0      \\
        \vdots & \vdots  & \vdots  & \vdots  & \vdots  & \vdots  & \vdots  & \vdots  & \vdots  & \vdots    & \vdots & \ddots & \vdots \\
        0      & 0       & 0       & 0       & 0       & 0       & 0       & 0       & 0       & 0         & 0      & \cdots & s_n
    \end{pmatrix}
\end{equation}

We impose an additional set of accuracy conditions, namely that the divergence of $f(r) = r^3$ also be exact everywhere except near the outer boundary.

The detailed accuracy conditions are
\begin{align}
    (\mathrm{D} \, r)_i   & = p+1 \quad \quad \quad i=1 \ldots N   , \\
    (\mathrm{D} \, r^3)_i & = (p+3)\, r^2 \quad i=1 \ldots N-N_c ,   \\
    (\mathrm{D} \, r^5)_i & = (p+5)\, r^4 \quad i=1 \ldots 3 ,       \\
    \int \mathbf{1} \, dx & = \sum_i s_i = \frac{R^{p+1}}{p+1},
\end{align}
where $N_c$ is again the boundary closure width. As above, the number $3$ in the third condition is chosen to ensure that the number of linearly independent equations matches the number of unknowns. We construct and solve a linear system for the values $s_i$ and $v_{ij}$ in the same way as in the previous subsection. We again find that $s_i=r^p_i h_i$ outside of the block near the origin.

Using the Cartesian operators from Mattsson et. al. \cite{Mattsson2004a} (Note one can use various cartesian operators to start from, such as the ones in \cite{Diener2007, Mattsson2004b, Mattsson2012, Mattsson2008}), we start with the norm $H$
\begin{align}
    \mathrm{H} & = \operatorname{diag}\!\left(\frac{1}{2},1,\dots,1, \frac{43801}{43200}, \frac{7877}{8640}, \frac{5359}{4320}, \frac{2711}{4320}, \frac{12013}{8640}, \frac{13649}{43200} \right),
\end{align}
the scalar norm $S$ is
\begin{align}
    \mathrm{S} = \operatorname{diag}\!\Biggl(
     & \dfrac{1430827147971}{4664636476456},
    \dfrac{42904697232509}{51311001241016},
    \dfrac{1539566352701}{424057861496},
    \dfrac{474102956851443}{51311001241016},
    \dfrac{37409168459741}{2332318238228},
    \dfrac{1279079894309069}{51311001241016},\notag \\
     & \dfrac{168033579790977}{4664636476456},
    \dfrac{1257018716149729}{25655500620508},
    \dfrac{298538403436579}{4664636476456},
    \dfrac{4156190224075161}{51311001241016},\notag \\
     & 10^2,11^2,\dots,27^2,\notag                  \\
     & \dfrac{1095025}{1728},
    \dfrac{1331213}{2160},
    \dfrac{144693}{160},
    \dfrac{132839}{270},
    \dfrac{10102933}{8640},
    \dfrac{13649}{48}
    \Biggr).
\end{align}

The diagonal entries of the vector norm $V$ (again for $p=2$) are then:
\begin{align}
    \mathrm{V}_{\rm diag} = \operatorname{diag}\Biggl(
     & 1,
    \frac{69612410558321}{51311001241016},
    \frac{3923059854503}{1509147095324},
    \frac{3014861355653139}{359177008687112},
    \frac{11554098239601139}{718354017374224},
    \frac{29105204758165}{1166159119114},
    \notag                                     \\
     & \frac{461864744704269}{12827750310254},
    7^2, 8^2, \dots, 23^2, 24^2,
    \frac{1095025}{1728},
    \frac{1331213}{2160},
    \frac{144693}{160},
    \frac{132839}{270},
    \frac{10102933}{8640},
    \frac{13649}{48}
    \Biggr),
\end{align}
and the off diagonal entries of $V$ are:
\begin{align}
    \mathrm{V}_{2,3} & = \dfrac{4996740529431}{6413875155127}    \\
    \mathrm{V}_{2,4} & = -\dfrac{13306004610507}{51311001241016} \\
    \mathrm{V}_{3,4} & = \dfrac{949724456067}{1166159119114}     \\
    \mathrm{V}_{3,5} & = -\dfrac{11638692514107}{51311001241016} \\
    \mathrm{V}_{4,5} & = \dfrac{11797110150741}{359177008687112} \\
    \mathrm{V}_{5,6} & = \dfrac{239755863585}{4664636476456}     \\
    \mathrm{V}_{6,7} & = -\dfrac{70992217935}{12827750310254}
\end{align}

\subsection{Comparison to plain staggered operators}
\label{sec:staggered}

Instead of constructing derivative operators that include a grid point on the origin, one can also construct staggered operators. We consider here the ``plain'' case, i.e. defining such operators without imposing any additional accuracy constraints. We show that such operators, if they satisfy the SBP property, are stable, but have significantly larger errors than the operators we defined above.

The points of a staggered grid lie at $r_i = (i-\frac{1}{2})\,h$. Assuming a Cartesian norm operator $H$, we can define scalar and vectors norms via
\begin{eqnarray}
    S = V &=& r_i^p\,H
\end{eqnarray}
i.e. by scaling the Cartesian norm in a straightforward manner. Since all $r_i>0$ this defines a norm.

The divergence is then defined as above via
\begin{eqnarray}
    \label{eqn:sbp-div-staggered}
    D = S^{-1} \left( B - G^T V \right)
\end{eqnarray}
This leads to a set of SBP operators.

One might be tempted to define the divergence directly via $D_\mathrm{bad} = G + p/r$. While well-defined, such operators will in general not satisfy the SBP property, and they lead in general to unstable (growing) modes. The problem is that the identity $r^{-p}\, \partial_r\, r^p = p/r$ holds in the continuum, but not for the discrete derivative operator. This is not a suitable way to discretize the divergence operator.

However, it \emph{is} possible to split the divergence into a partial derivative plus a discrete Christoffel symbol while retaining the SBP property, leading to a stable setup. For a Cartesian SBP operator we have
\begin{eqnarray}
    H G + (H G)^T &=& B\,.
\end{eqnarray}
We introduce $R = \mathrm{diag}(r_i^p)$ as operator that scales with the effective volume element.
We multiply by $R$ from the right, and insert $R R^{-1}$ into the first term. We obtain
\begin{eqnarray}
    H R R^{-1} G R + (R H G)^T &=& B R\,.
\end{eqnarray}
We now define $S := H R$, $V := R H$, $D := R^{-1} G R$, and rescale $B$. We thus find a new set of SBP operators that have been rescaled by $R$:
\begin{eqnarray}
    S D + (V G)^T &=& B\,.
\end{eqnarray}
That is, $D = R^{-1} G R$ is a consistent rescaling that defines a divergence that satisfies SBP. We can trivially rewrite this as
\begin{eqnarray}
    D &=& G + \Gamma \\
    \Gamma &=& = R^{-1} G R - G\, .
\end{eqnarray}
Of course, this is just a rewritten version (\ref{eqn:sbp-div-staggered}) with exactly the same properties, apart from floating-point accuracy. This ansatz defines the Christoffel symbol $\Gamma$ in such a way that it exactly complements the discrete partial derivative $G$. $\Gamma$ is a linear operator that discretizes the operation ``multiply by $(p/r)$'', which is indeed linear operation.

\begin{figure}[H]
    \centering
    \includegraphics[width=0.49\textwidth]{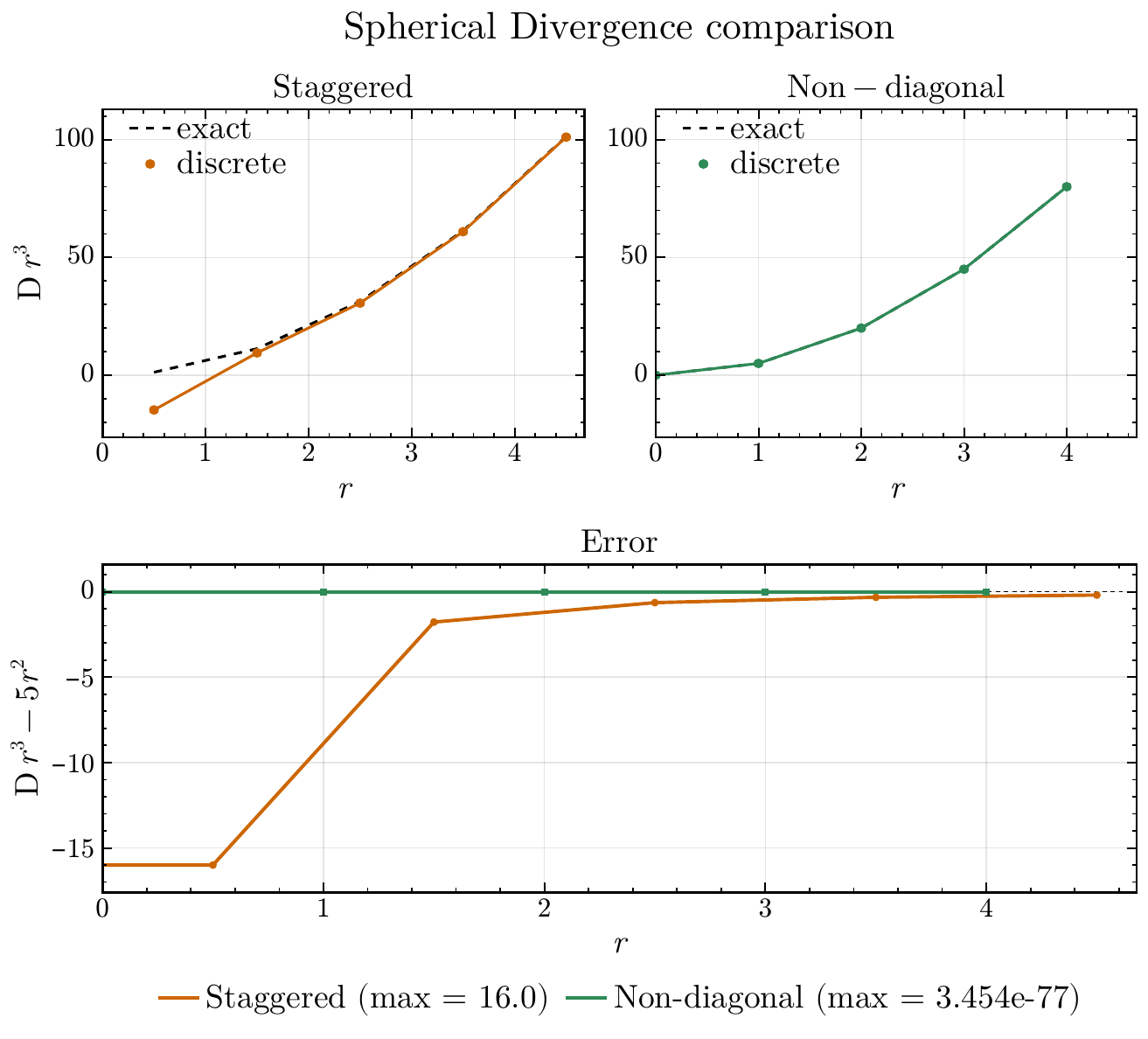}
    \includegraphics[width=0.49\textwidth]{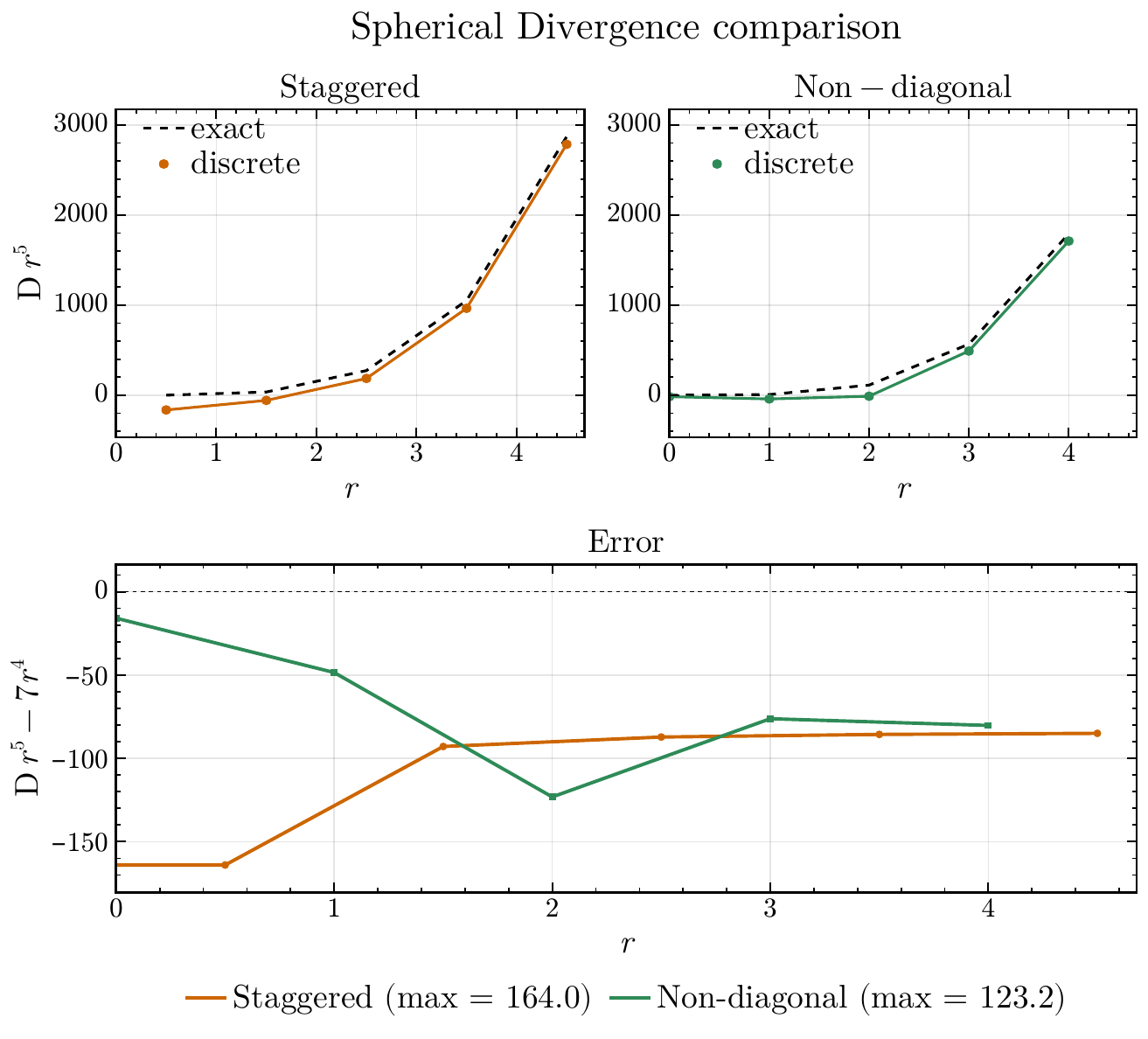}
    \caption{Discretization error when applying 4th-order discrete divergence operator to $f(r) = r^3$ and $f(r) = r^5$, for spherical symmetry ($p=2$). We compare a plain staggered operator to our non-diagonal operator. Our operator is exact for $r^3$ near the origin, and has a much smaller error near the origin for $r^5$.}
    \label{fig:div-comparison-o4}
\end{figure}

\begin{figure}[H]
    \centering
    \includegraphics[width=0.49\textwidth]{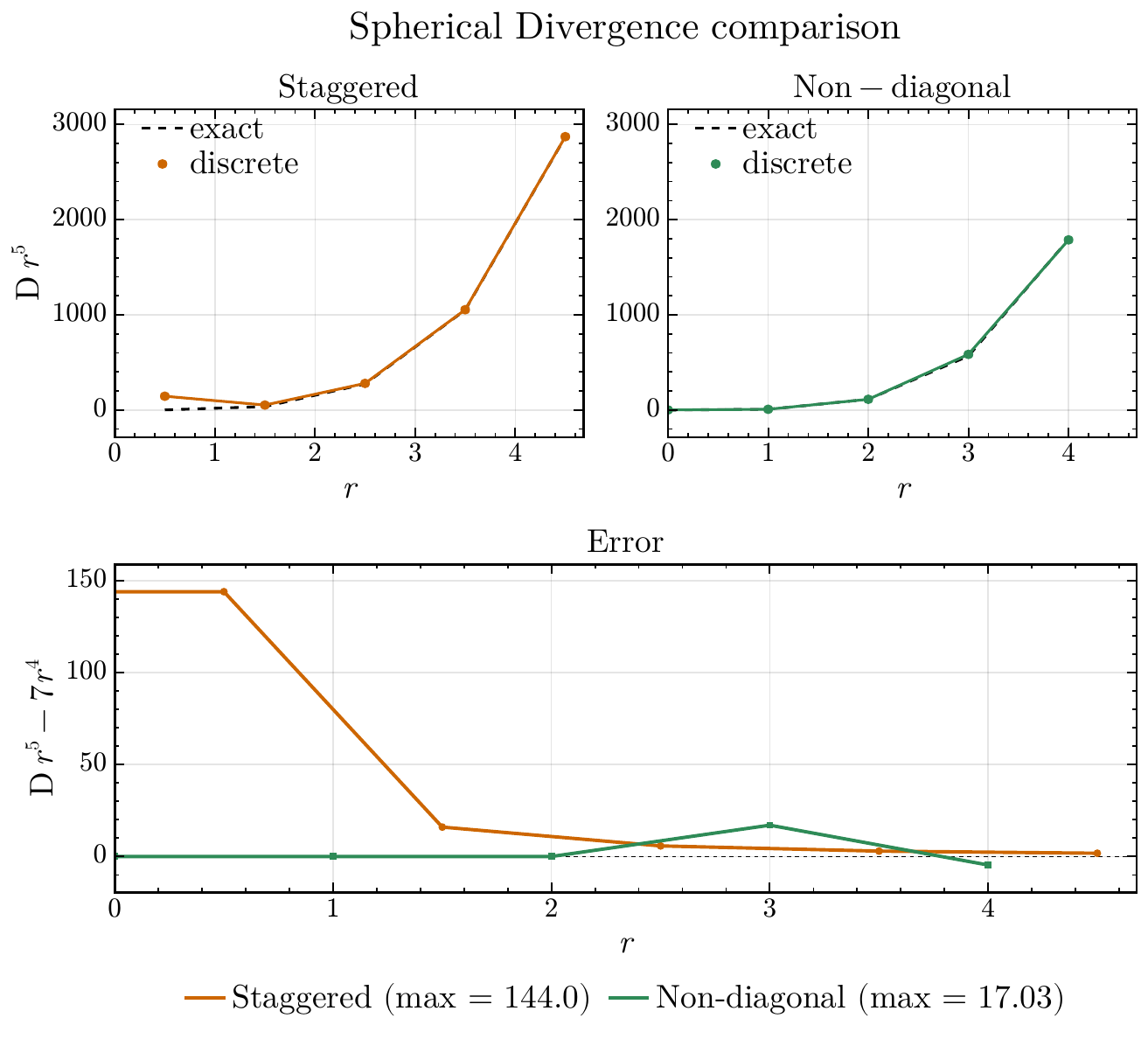}
    \includegraphics[width=0.49\textwidth]{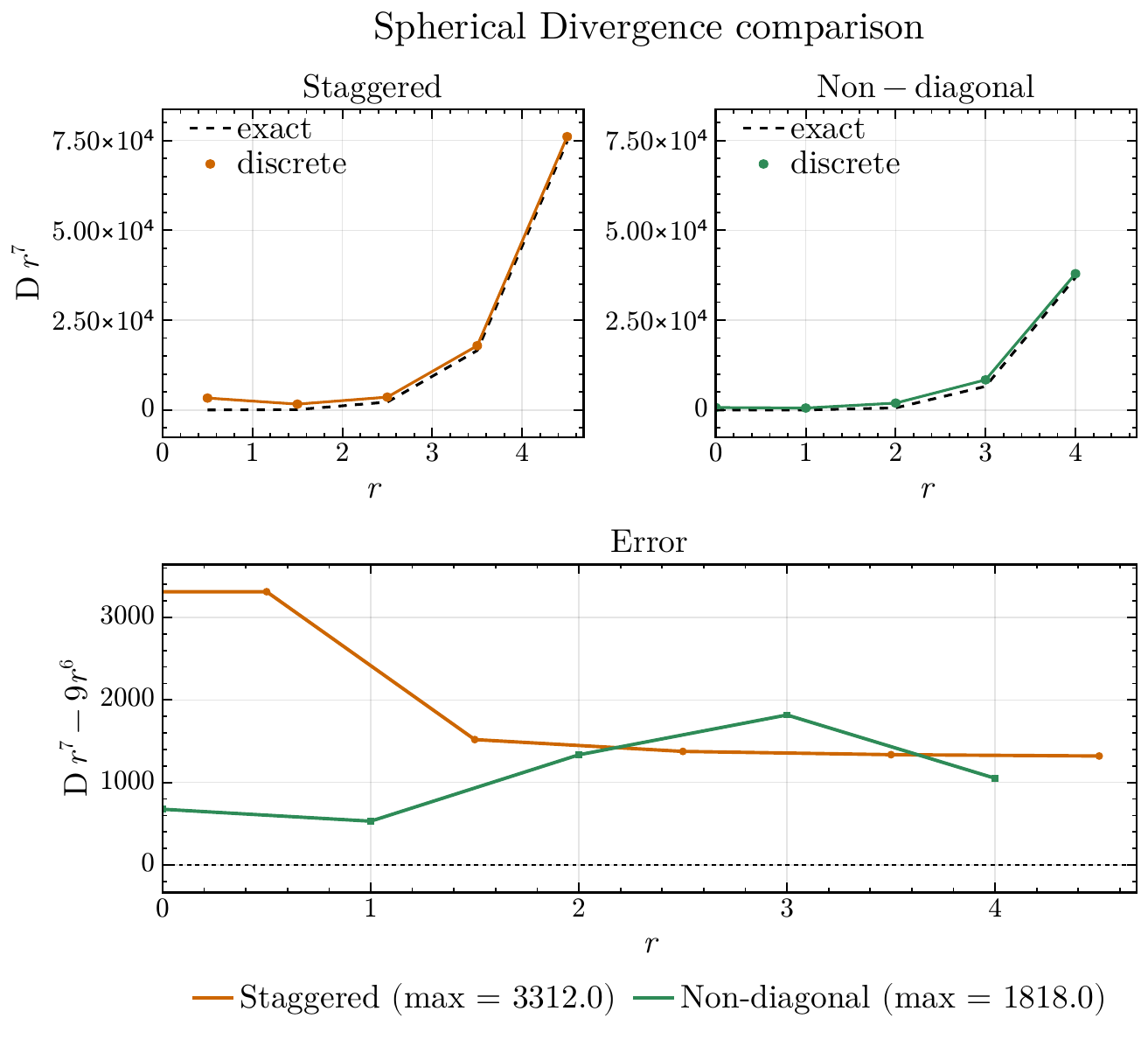}
    \caption{Discretization error when applying 6th-order discrete divergence operator to $f(r) = r^5$ and $f(r) = r^7$, for spherical symmetry ($p=2$). We compare a plain staggered operator to our non-diagonal operator. Our operator is exact for $r^5$ near the origin, and has a significantly smaller error near the origin for $r^7$.}
    \label{fig:div-comparison-o6}
\end{figure}

To compare this plain (without imposing additional accuracy conditions) staggered divergence to the operators we constructed in Equation \ref{eqn:sbp-div-staggered}, we apply these operators to the functions $f(r)=r^3$, $f(r)=r^5$, and $f(r)=r^7$, and compare the discretization errors. See Figures \ref{fig:div-comparison-o4} and \ref{fig:div-comparison-o6}. Recall that a discrete divergence loses, a priori, $p$ degrees of exactness compared to the discrete gradient since the divergence implicitly divides by $r^p$, as discussed in Section \ref{sec:order-and-degree}. We also compare the spectral radius of these operators, which affects the CFL factor, in the next section.

\section{Spectral Stability}

In the section above we discussed the stability of the semi-discrete system, i.e. we assumed a discretization in space while keeping the time direction continuous. In this section we comment on the stability of a fully discrete system, assuming that one uses an ODE integrator for time stepping.

The stability of a numerical algorithm solving the wave equation on the spectral properties of its spatial discretization. Its eigenvalues scaled by the timestep, $\lambda \cdot \Delta t$, need to lie within the stability region of the time integrator to avoid growing modes \cite{Hairer1996}. This imposes constraints on the eigenvalues of the spatial operators. (The eigenvalues will in general be complex numbers.) For example, the eigenvalues cannot have a positive real part, since this corresponds to a growing mode. Moreover, eigenvalues with a large magnitude need to be scaled by a small enough timestep to ensure they fall into the stability region of the time integrator. Eigenvalues with a large magnitude make an operator \emph{stiff}, i.e. they impose a restriction on the allowed time step size.

The continuous Laplace operator is symmetric and negative semi-definite, guaranteeing purely real, non-positive eigenvalues. In the discrete setting, constructing the Laplacian by multiplying two first-derivative SBP operators ($\mathrm{L}_{\mathrm{lap}} = \mathrm{D}\cdot\mathrm{G}$) places certain demands on the operators that we discuss below.

Transforming the second-order wave equation into a system of first-order equations changes the spatial discretization into a skew-symmetric block structure:
\begin{equation}
    \mathrm{L}_{\mathrm{block}} = \begin{bmatrix} 0 & \mathrm{G} \\ \mathrm{D} & 0 \end{bmatrix}\, .
    \label{eq:block_operator}
\end{equation}
Energy stability relies on the SBP property of the derivative operators. The eigenvalues of this block matrix are purely imaginary, representing undamped physical oscillation and propagation of numerical waves.

Applying boundary conditions modifies the real components of the block operator's spectrum.
\begin{itemize}
    \item \textbf{Reflective (energy-conserving) boundaries:} A properly formulated reflective boundary condition  preserves the discrete skew-symmetry of the block matrix. The real parts of the eigenvalues remain bounded at machine zero ($|\Re(\lambda)| \approx 10^{-14}$) and prevent unphysical growth or decay. The imaginary eigenvalue pairs with the largest magnitude define the highest resolvable frequencies and lead to the temporal stability bounds for explicit time integrators.

    \item \textbf{Radiative (energy-dissipating) boundaries:} To model an open domain, the radiative boundaries (e.g. applied as Simultaneous Approximation Terms (SAT)) break the global skew-symmetry by introducing a negative-definite penalty. This shifts the spectrum to the left half of the complex plane. While $\max \Re(\lambda)$ remains zero and guarantees global stability, $\min \Re(\lambda)$ becomes negative, depending on the rate of energy dissipation as the wave exits the computational domain.
\end{itemize}

To avoid numerical blow-up, the time step size $\Delta t$ must be constrained so that the scaled eigenvalues $\lambda \cdot \Delta t$ fall within the stability region of the time integrator. For a standard explicit scheme such as the classic fourth-order Runge-Kutta (\texttt{RK4}) method \cite{Shampine2005, Hairer1993}, the stability region extends along the imaginary axis up to approximately $\pm 2.82i$. Therefore, the stable time step sizes are restricted to
\begin{equation}
    \Delta t \lesssim \frac{2.82}{\max |\lambda|}
\end{equation}
The 8th-order accurate \texttt{DP8} method by Dormand \& Prince \cite{Hairer1993} that we use in the next section has a larger stability region with size $\sim 5.9$, allowing either larger time step sizes, or allowing using operators with a larger spectral radius.

We compare the largest eigenvalues and spectral radius of several discrete derivative operators in Table \ref{tab:spectral-radii}, where we also examine the staggered operators described in section \ref{sec:staggered} above. As the table shows, the SBP operators are stable, and radiative boundaries dissipate energy, as expected.
As described in the previous section, our new SBP operators have an increased accuracy near the origin, as prescribed in our accuracy conditions.

The staggered non-SBP operators we introduced in the previous section are unstable. This is not a CFL instability -- there are one or more growing modes in the discretization, independent of the time discretization. We did not investigate whether adding artificial dissipation can stabilize the system.

\begin{table}[H]
    \centering
    \small
    \setlength{\tabcolsep}{5pt}
    \begin{tabular}{l | l |
            S[table-format=1.2]
            S[table-format=-1.2]
            S[table-format=1.2] |
            S[table-format=1.2]
            S[table-format=-1.2]
            S[table-format=1.2]}
              &                   & \multicolumn{3}{c|}{reflecting (Dirichlet) boundaries} & \multicolumn{3}{c}{ radiative (SAT) boundaries}                                                                                             \\
        Order & Setup             & {$\max \Re(\lambda)$}                                  & {$\min \Re(\lambda)$}                           & {$|\rho(\lambda)|$} & {$\max \Re(\lambda)$} & {$\min \Re(\lambda)$} & {$|\rho(\lambda)|$} \\\hline
        4th   & our operators     & 0                                                      & 0                                               & 1.94                & 0                     & -0.293                & 1.851               \\\
              & SBP staggered     & 0                                                      & 0                                               & 2.0                 & 0                     & -0.291                & 2.0                 \\\
              & non-SBP staggered & 0.025                                                  & -0.025                                          & 1.98                & 0.024                 & -0.233                & 1.4                 \\\hline
        6th   & our operators     & 0                                                      & 0                                               & 2.47                & 0                     & -0.331                & 1.981               \\\
              & SBP staggered     & 0                                                      & 0                                               & 2.47                & 0                     & -0.329                & 2.20                \\\
              & non-SBP staggered & 0.028                                                  & -0.028                                          & 2.44                & 0.036                 & -0.260                & 1.930
    \end{tabular}
    \caption{Largest real eigenvalues and spectral radii $\rho$ for several derivative operators using the cartesian operators from \cite{Mattsson2004a}, assuming the first-order discretization in (\ref{eq:block_operator}). As discussed in the main text, a larger spectral radius requires smaller time step sizes. Eigenvalues with a positive real part indicate a growing mode (an instability), a negative real part indicates damped modes (possibly from outgoing boundary conditions). All SBP operators are stable, and the non-SBP staggered operators are unstable (they have a growing mode). The radiative boundaries dissipate energy, as expected. Other cartesian operators produced the same qualitative results.}
    \label{tab:spectral-radii}
\end{table}

\section{Numerical evolutions}
\label{sec:numerical}

We demonstrate the higher-order stencils by solving the scalar wave equation in spherical symmetry. This corresponds to setting $p=2$. We set up a Gaussian wave packet that starts away from the origin, moves inwards, ``reflects'' at the origin (or moves through the origin, depending on point of view), and the expands outwards again.

The general solution for this PDE is given by
\begin{eqnarray}
    \Phi(t,r) &=& \frac{g(t+r)-g(t-r)}{r}
\end{eqnarray}
for an arbitrary differentiable function $g$ which depends only on a single argument. This ansatz satisfies the scalar wave equation in spherical symmetry, and by inspection it is clear that$\Phi$ does not have any singularities at the origin. Numerically, l'Hôptial's rule needs to be used at the origin.

We choose $g$ to be a Gaussian
\begin{eqnarray}
    g(u) &=& A \exp\!\left(-\frac{(u-r_c)^2}{d^2}\right)\, .
\end{eqnarray}
with amplitude $A=1$, width $d=2$, and initial position $r_c = 10$ for a domain $[0,40]$. We use reflecting boundary conditions at the outer boundary $r=R$ (although we are not interested in the behavior near the outer boundary here -- we want to demonstrate convergence and accuracy near the origin).

Time integration is performed with Hairer's 8/5/3 adaption of the Dormand-Prince Runge-Kutta method \texttt{DP8} \cite{Hairer1993} from \texttt{OrdinaryDiffEq.jl} \cite{Rackauckas2017}, used with a fixed time-step \(\Delta t\) and adaptive stepping disabled. The CFL factor is $\mathrm{cfl}=0.5$.

\begin{figure}[H]
    \centering
    \includegraphics[width=1.0\textwidth]{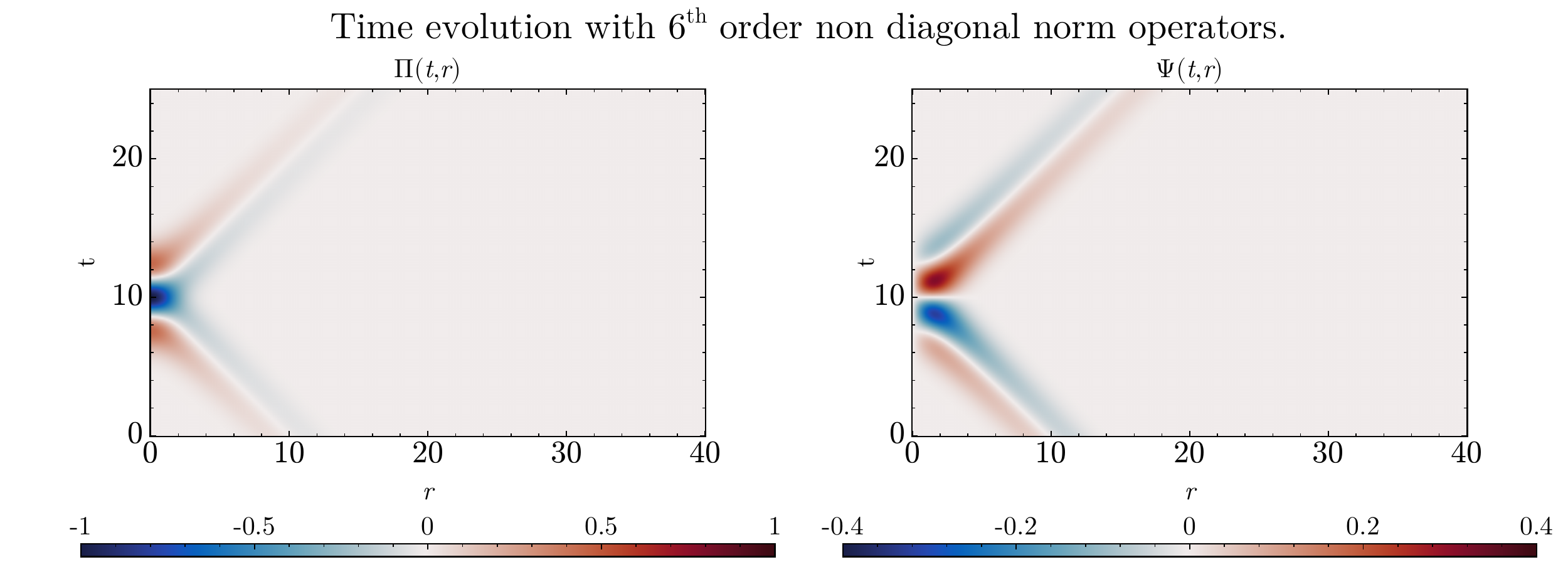}
    \caption{Time evolution for the 6th-order non-diagonal SBP operator in spherical symmetry ($p=2$), shown at resolution $h=1/16$ for $0 \le t \le 25$. The corresponding time evolutions for the other operators are visually indistinguishable.}
    \label{fig:wave-time-evolution-non_diagonal-order6-reflective}
\end{figure}

We show in Figure \ref{fig:wave-time-evolution-non_diagonal-order6-reflective} a spacetime plot which gives an overview over the evolution of this system. The wave packet starts at $r_c=10$, and then moves inwards as it amplitude increases like $1/r$. It interacts with the origin and then moves outwards again, its amplitude falling off as $1/r$. The overall evolution looks clean, although this is not a very strong statement given the coarse scale of the color map.

\subsection{Energy Conservation}

\begin{figure}[H]
    \centering
    \begin{subfigure}[t]{0.45\textwidth}
        \centering
        \includegraphics[width=\textwidth]{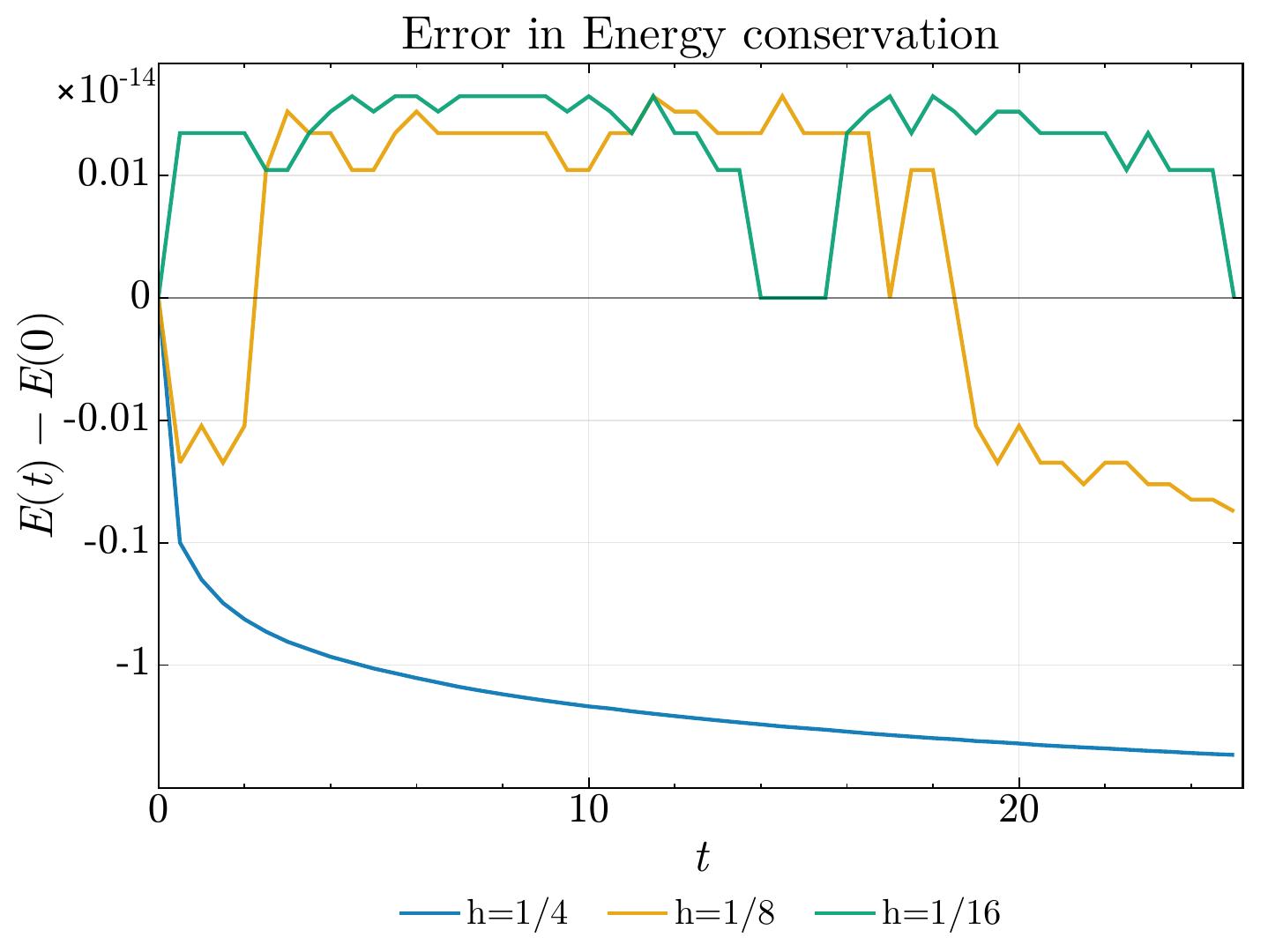}
        \caption{4th order non-diagonal operator}
    \end{subfigure}
    \hfill
    \begin{subfigure}[t]{0.45\textwidth}
        \centering
        \includegraphics[width=\textwidth]{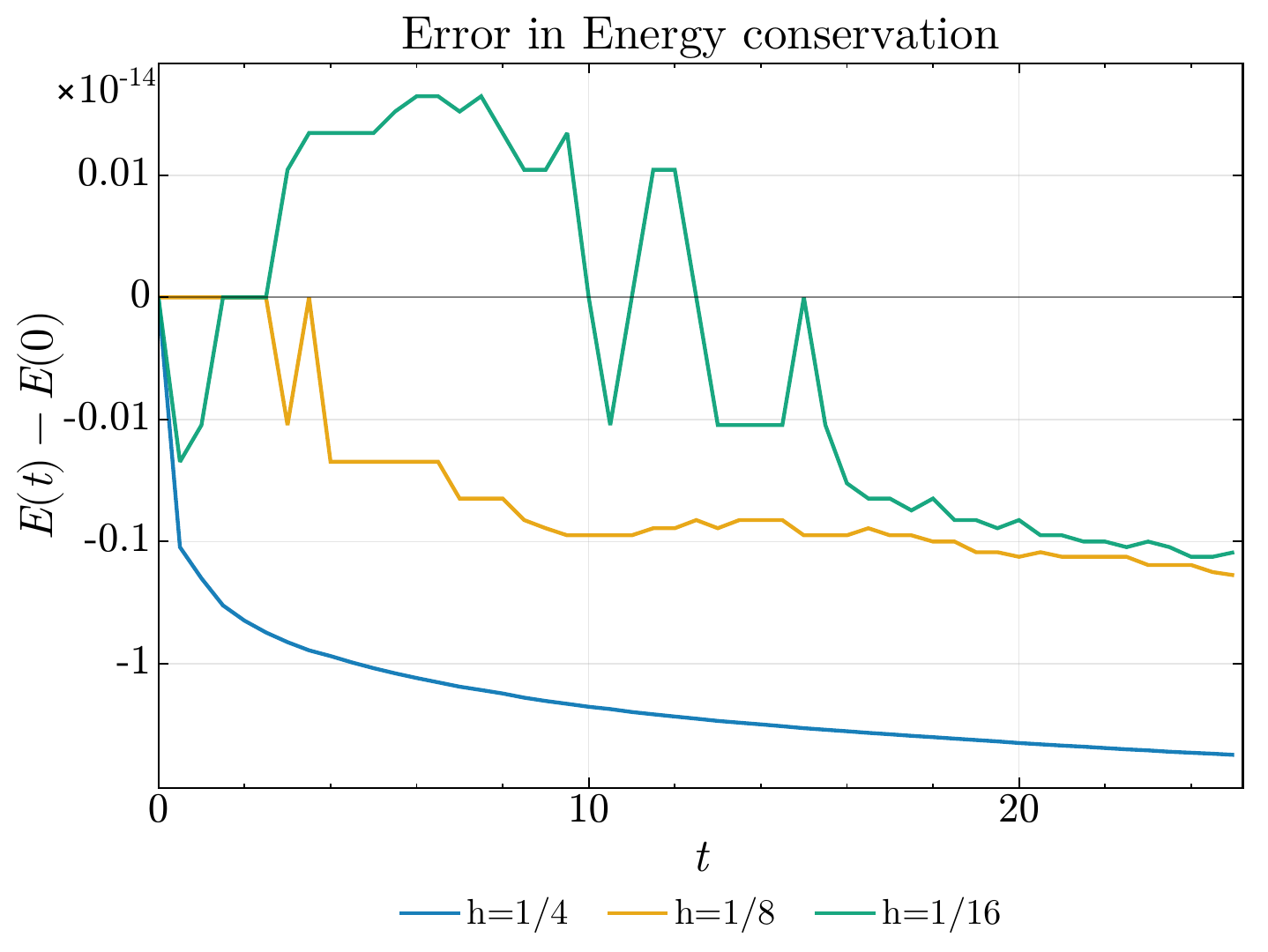}
        \caption{6th order non-diagonal operator}
    \end{subfigure}
    \caption{Error in energy conservation for the non-diagonal-norm SBP wave evolutions. The left and right subfigures show the 4th- and 6th-order operators using $p=2$ and $R=40$. Three resolutions are shown: $h=(1/4, 1/8, 1/16)$ over the time interval $0 \le t \le 25$, using a CFL factor of $1/2$. Note the graphs are scaled by $10^{-14}$, i.e. they are showing only floating-point round-off errors. The lowest resolution $h=1/4$ shows a slight decrease in the total energy over time. We confirmed (not shown here) that this can also cured by reducing the CFL factor.}
    \label{fig:wave-energy-non-diagonal}
\end{figure}

The discrete energy of the system is
\begin{equation}
    E(t) = \frac{1}{2}\left(\Pi S \Pi + \Psi V \Psi \right).
\end{equation}
We use reflecting boundary conditions, and thus energy must be conserved. We show in figure \ref{fig:wave-energy-non-diagonal} the error in energy conservation $E(t)-E(0)$ up to $t=25$. At this time the wave packet has reflected off the origin. The graphs show that energy is indeed conserved up to floating-point round-off errors, demonstrating the stability of our discretization.

\begin{figure}[H]
    \centering
    \begin{subfigure}[t]{0.45\textwidth}
        \centering
        \includegraphics[width=\textwidth]{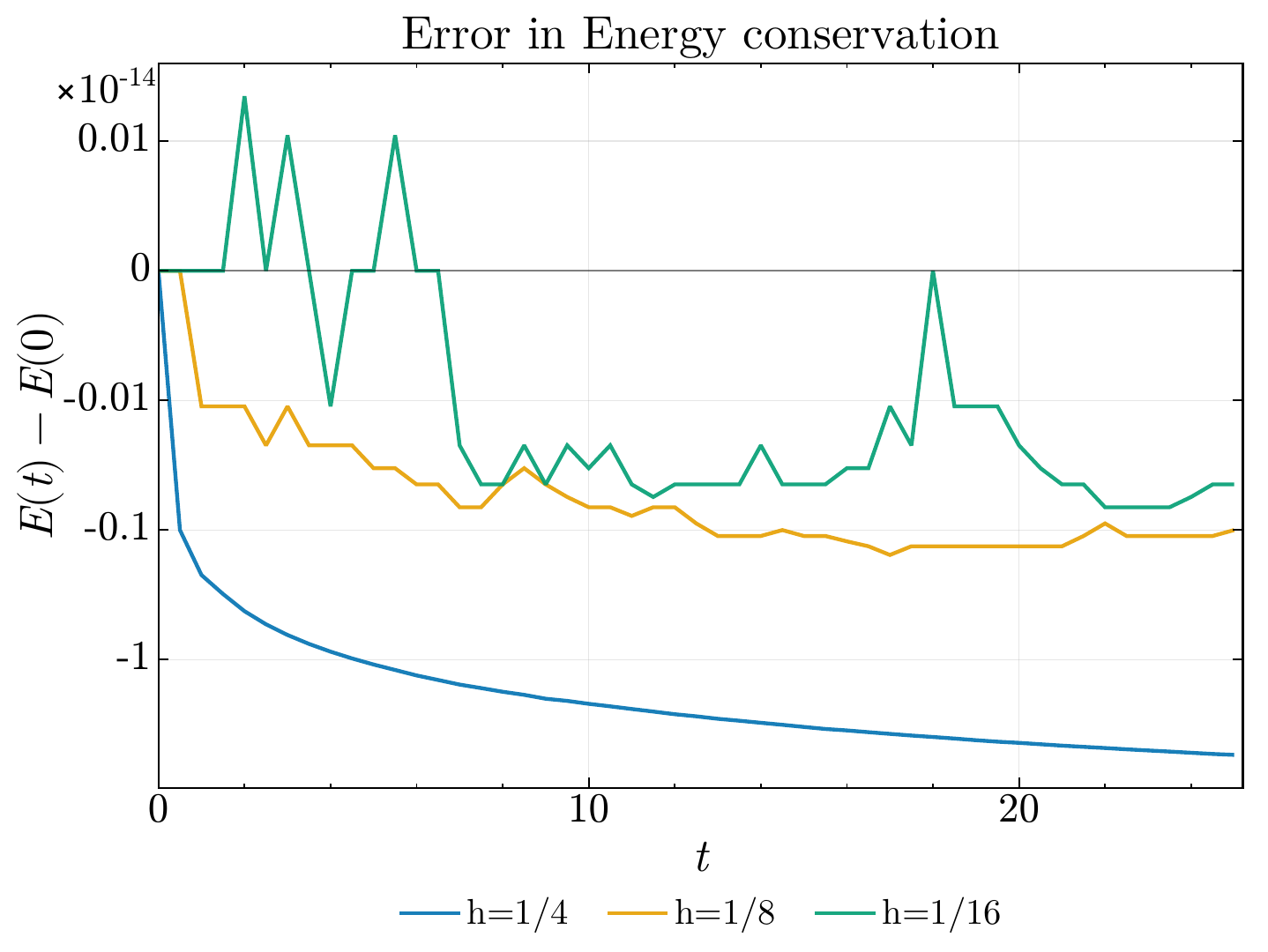}
        \caption{4th order staggered operator}
    \end{subfigure}
    \hfill
    \begin{subfigure}[t]{0.45\textwidth}
        \centering
        \includegraphics[width=\textwidth]{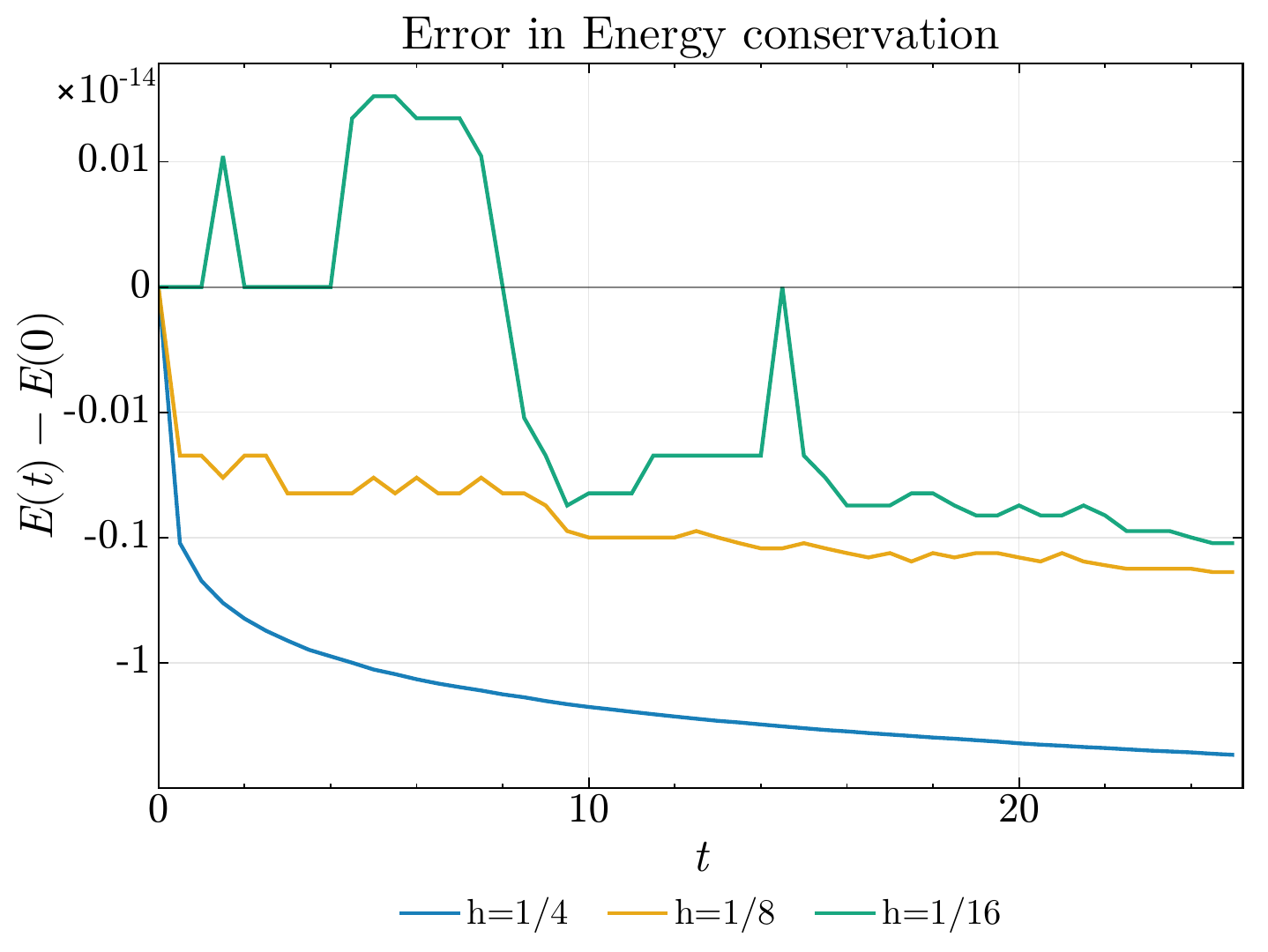}
        \caption{6th order staggered operator}
    \end{subfigure}
    \caption{Error in energy conservation for the staggered SBP wave evolutions. The left and right subfigures show the 4th- and 6th-order operators using $p=2$ and $R=39.5$. Three resolutions are shown: $h=(1/4, 1/8, 1/16)$ over the time interval $0 \le t \le 25$, using a CFL factor of $1/2$. Note the graphs are scaled by $10^{-14}$, i.e. they are showing only floating-point round-off errors. The lowest resolution $h=1/4$ shows a slight decrease in the total energy over time. We confirmed (not shown here) that this can also cured by reducing the CFL factor.}
    \label{fig:wave-energy-staggered}
\end{figure}

\subsection{Accuracy and Convergence}

\begin{figure}[H]
    \centering
    \includegraphics[width=1.0\textwidth]{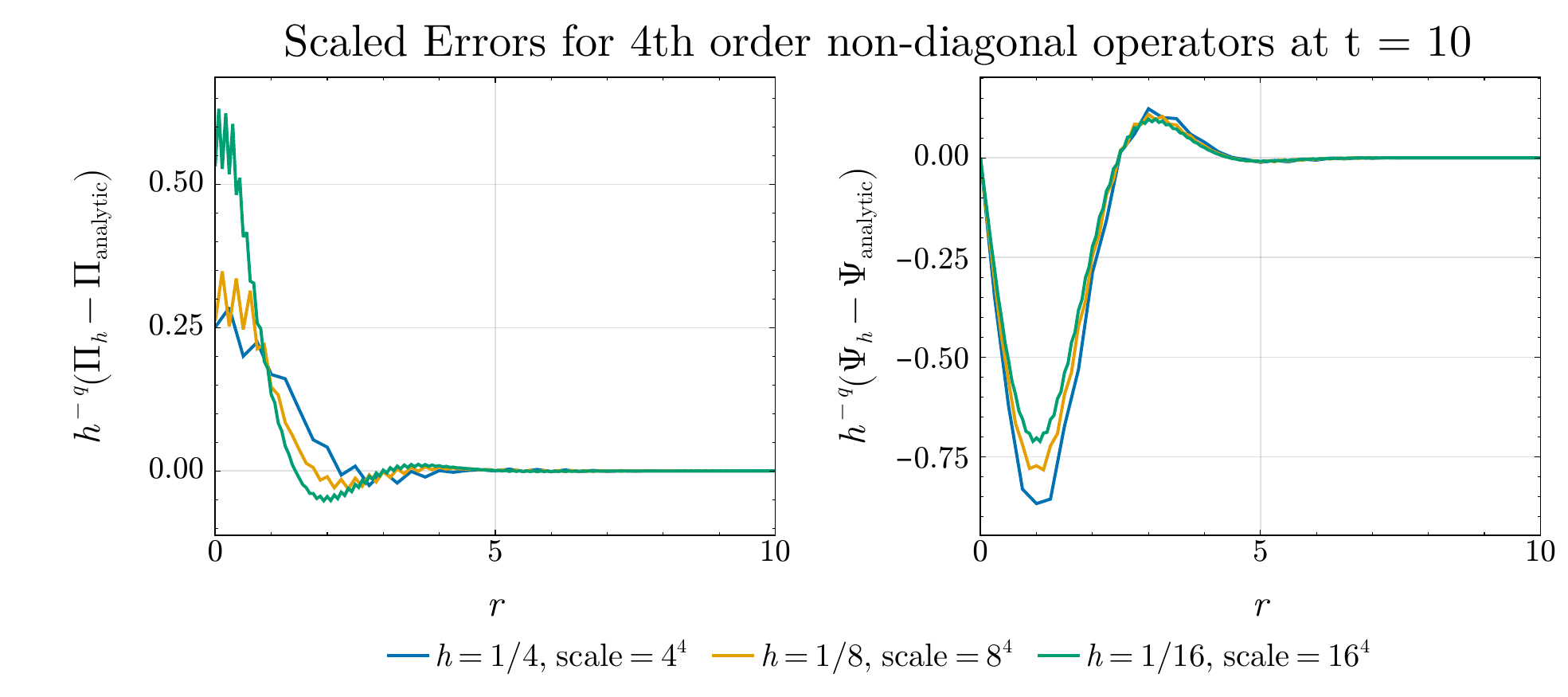}
    \includegraphics[width=1.0\textwidth]{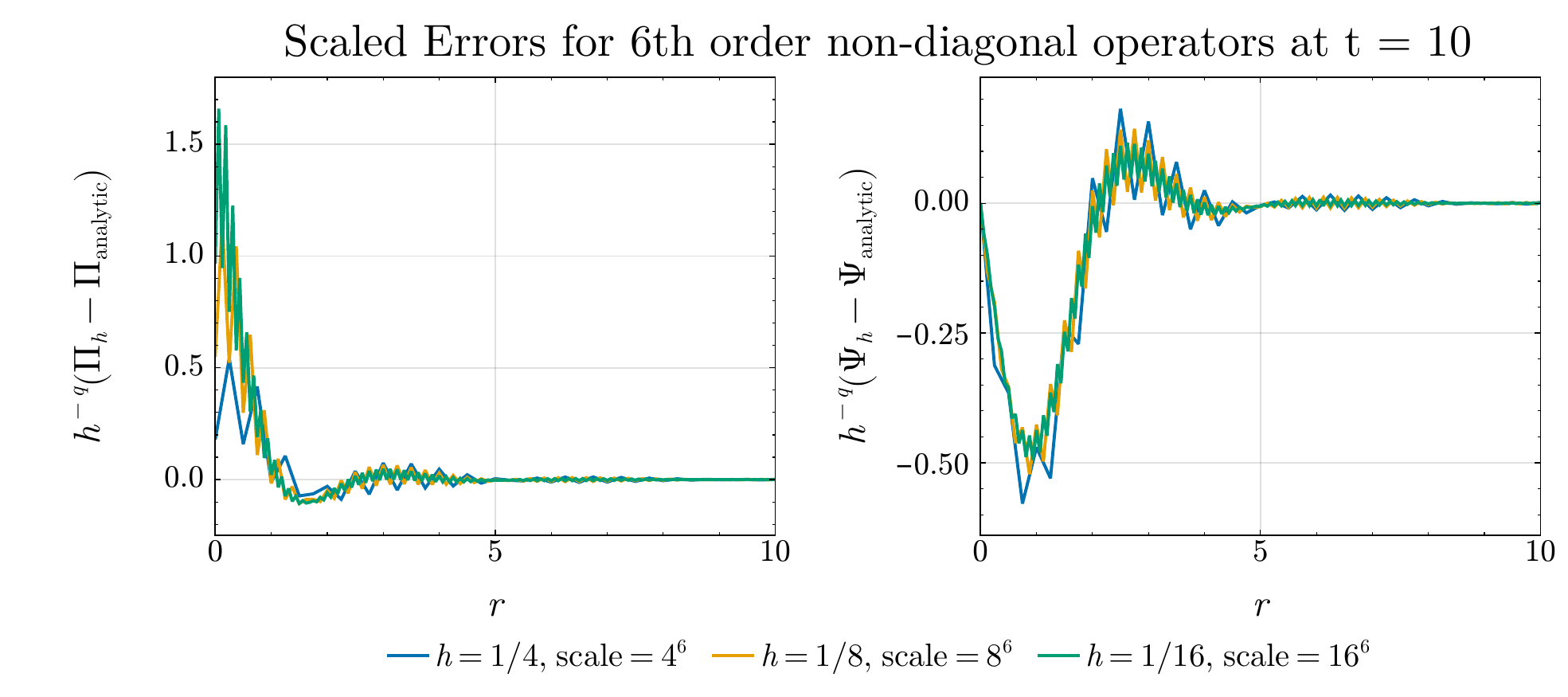}
    \caption{Convergence diagnostics for the 4th-order and 6th-order non-diagonal-norm SBP operators at $t=10$ for the resolutions $h=(1/4,1/8,1/16)$. Shows scaled errors (not the solution!), scaled by $h^{-4}$. This is the configuration $p=2$ on $r=[0,40]$. The noise amplitude for the highest resolution $h=1/16$ is at the level of $10^{-8}$ and converges away with increase resolution.}
    \label{fig:wave-convergence-banded-early}
\end{figure}

\begin{figure}[H]
    \centering
    \includegraphics[width=1.0\textwidth]{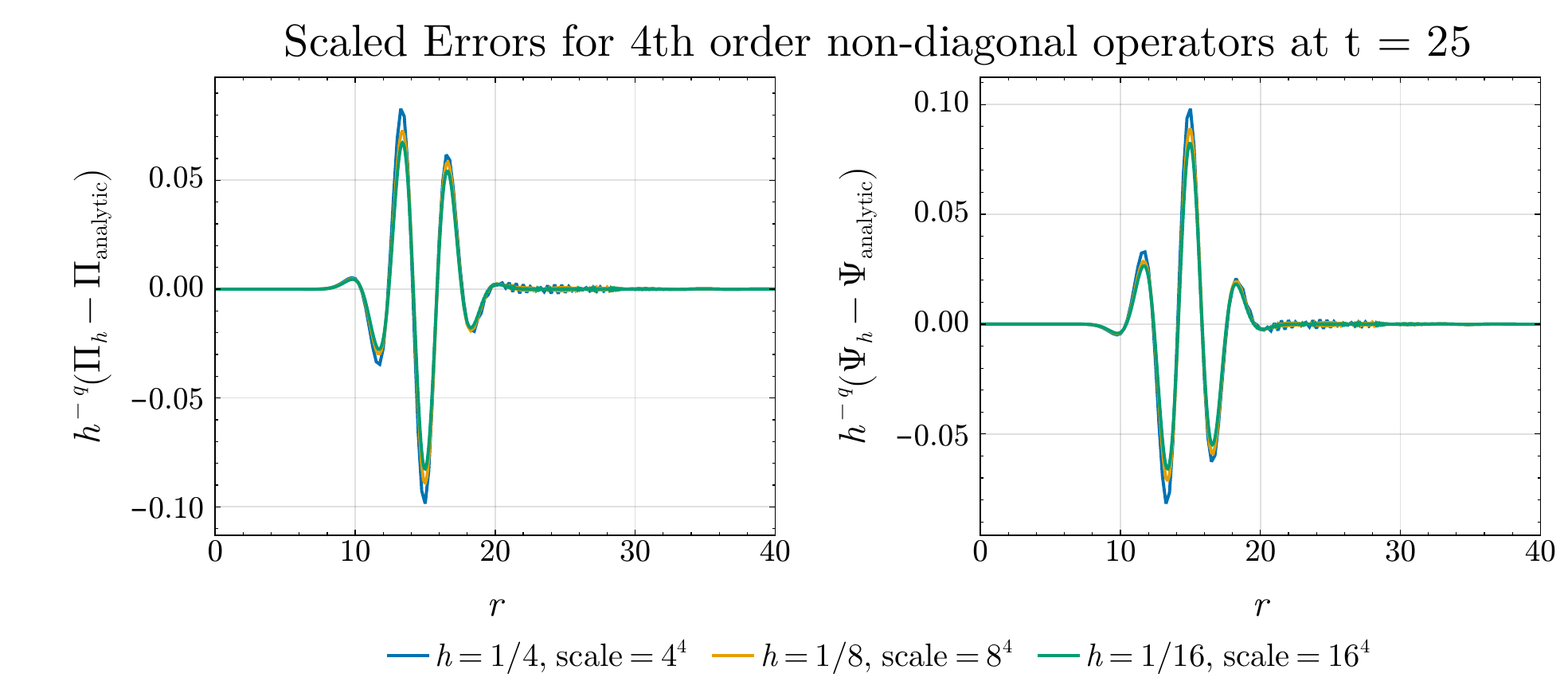}
    \includegraphics[width=1.0\textwidth]{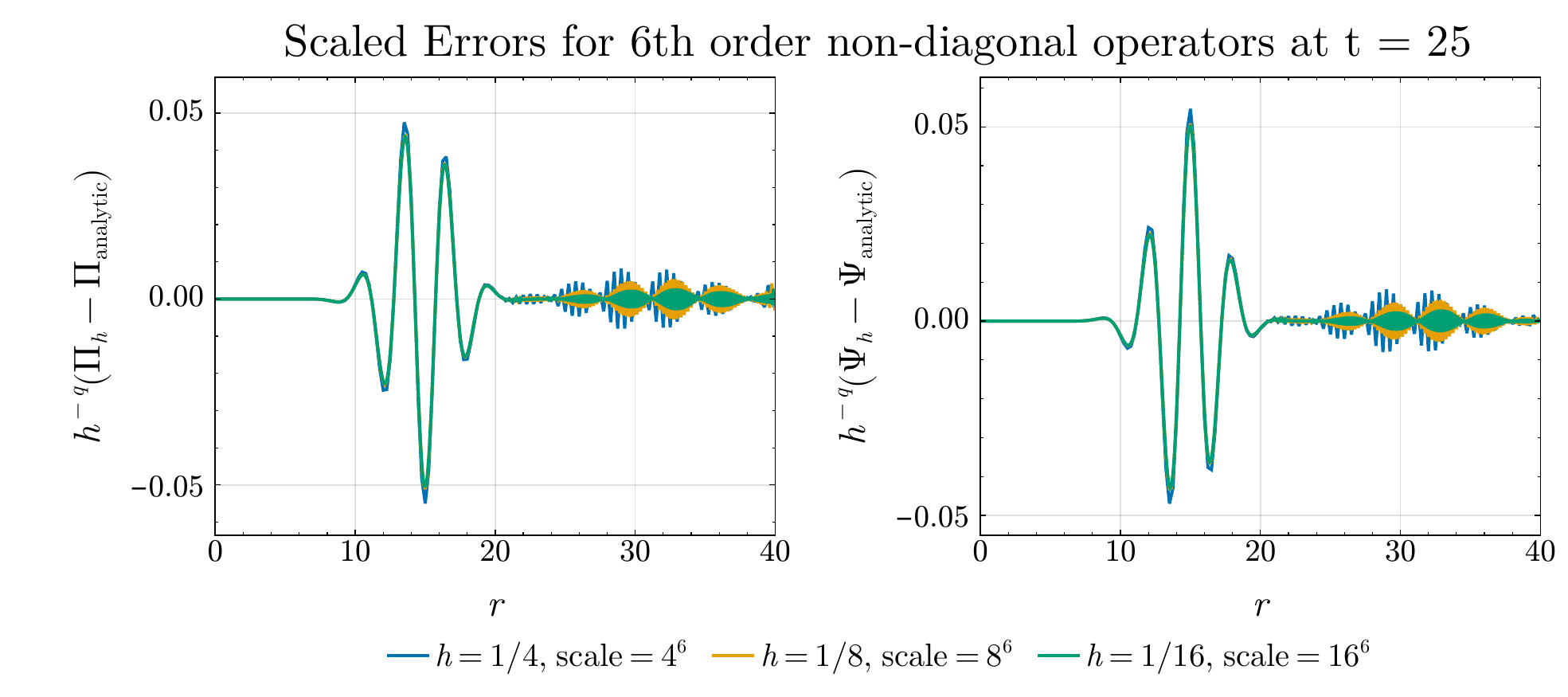}
    \caption{Convergence diagnostics for the 4th-order and 6th-order non-diagonal-norm SBP operator at $t=25$ for the resolutions $h=(1/4,1/8,1/16)$. Shows scaled errors (not the solution!), scaled by $h^{-4}$. This is the configuration $p=2$ on $r=[0,40]$. The noise amplitude for the highest resolution $h=1/16$ is at the level of $10^{-9}$ and converges away with increased resolution.}
    \label{fig:wave-convergence-banded-late}
\end{figure}

We investigate the accuracy and convergence of the operators by showing the scaled differences to the exact solution for different resolutions at $t=10$ and $t=25$. At $t=10$, the pulse is interacting with the origin. At $t=25$, the pulse has reflected off the origin and is moving outwards. Figures \ref{fig:wave-convergence-banded-early} and \ref{fig:wave-convergence-banded-late} shows the scaled errors at these times. The errors have been scaled according to the expected convergence order (4 and 6, respectively), and good convergence is indicated by scaled errors that lie on top of each other. At $t=10$, near the origin, the convergence is not quite clean, and a small amount of noise is visible.  At late times ($t=25$), convergence is clean again, and there is a small amount of noise visible that propagates faster than the pulse. Please consider that these figures show the error, not the solution, and that the high-resolution errors have been scaled by $16^6 \approx 1.6\cdot10^{-7}$, i.e. these errors are quite small. The figures also show that the noise converges away with increasing resolution.

\begin{figure}[H]
    \centering
    \includegraphics[width=1.0\textwidth]{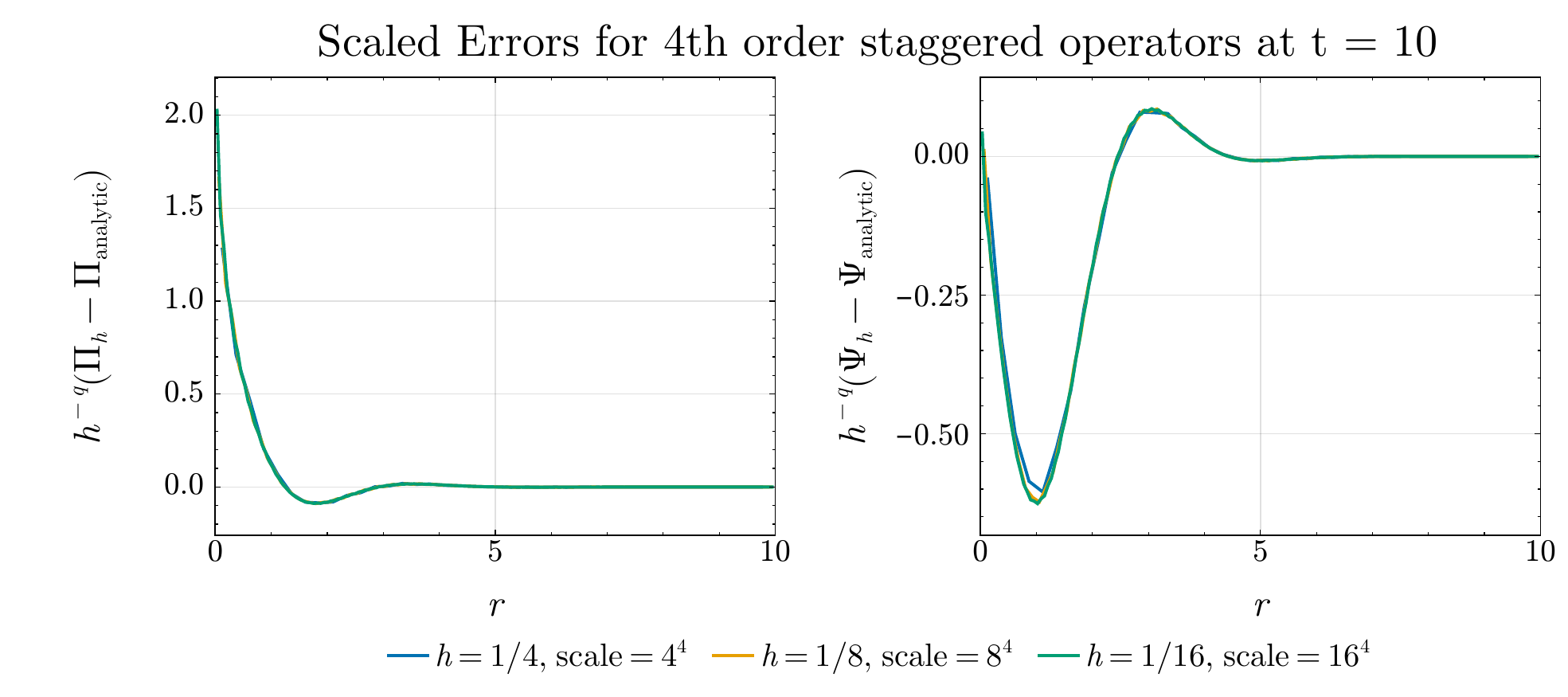}
    \includegraphics[width=1.0\textwidth]{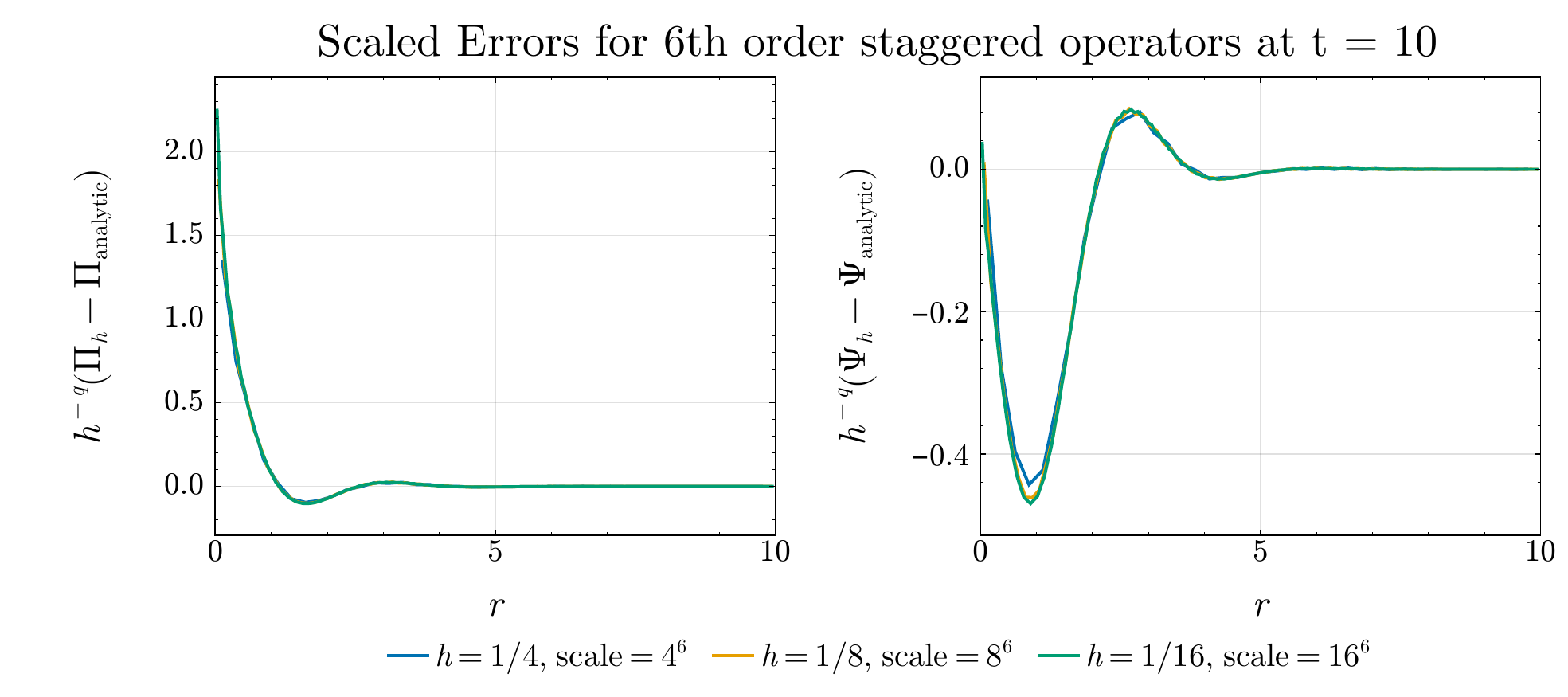}
    \caption{Convergence diagnostics for the 4th-order and 6th-order staggered SBP operators at $t=10$ for the resolutions $h=(1/4,1/8,1/16)$. Shows scaled errors (not the solution!), scaled by $h^{-4}$. This is the configuration $p=2$ on $r=[0,39.5]$. The noise amplitude for the highest resolution $h=1/16$ is at the level of $10^{-8}$ and converges away with increase resolution.}
    \label{fig:wave-convergence-staggered-early}
\end{figure}

\begin{figure}[H]
    \centering
    \includegraphics[width=1.0\textwidth]{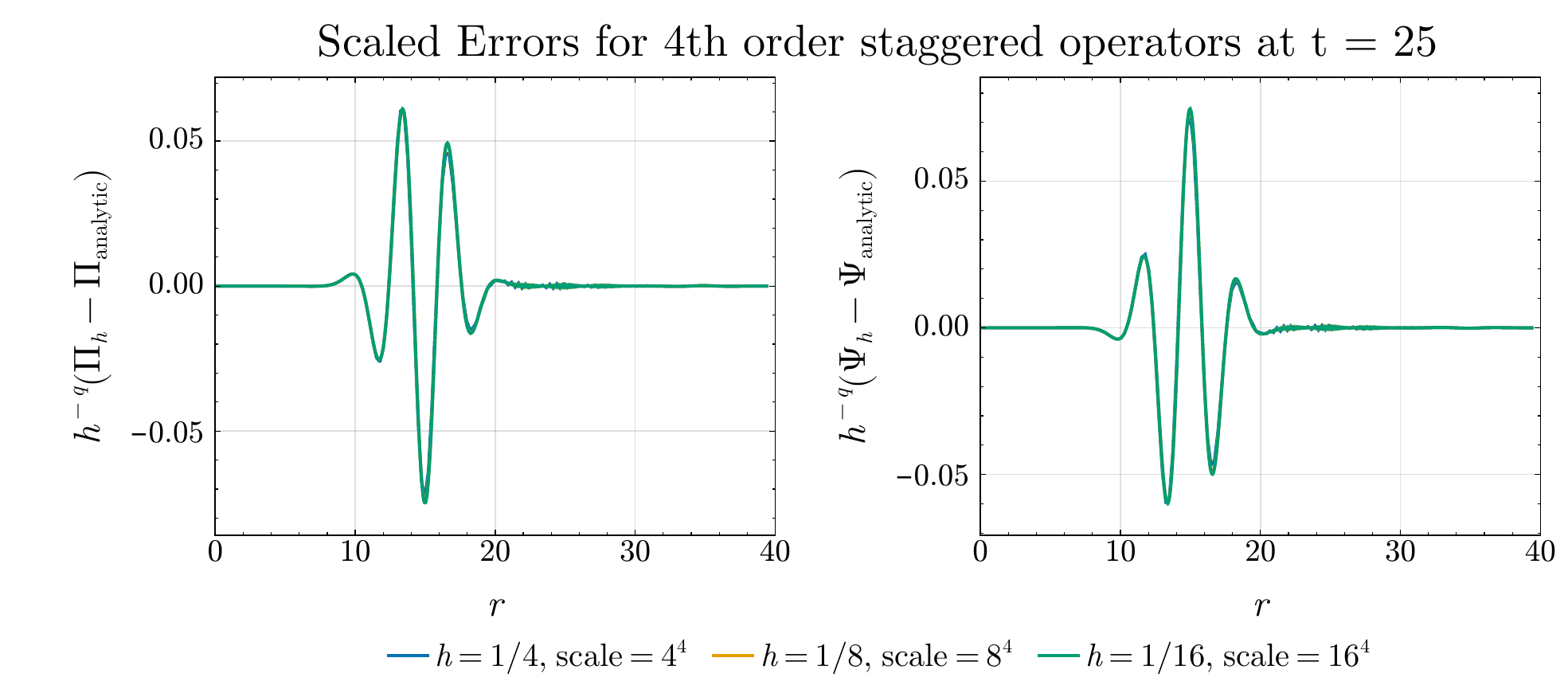}
    \includegraphics[width=1.0\textwidth]{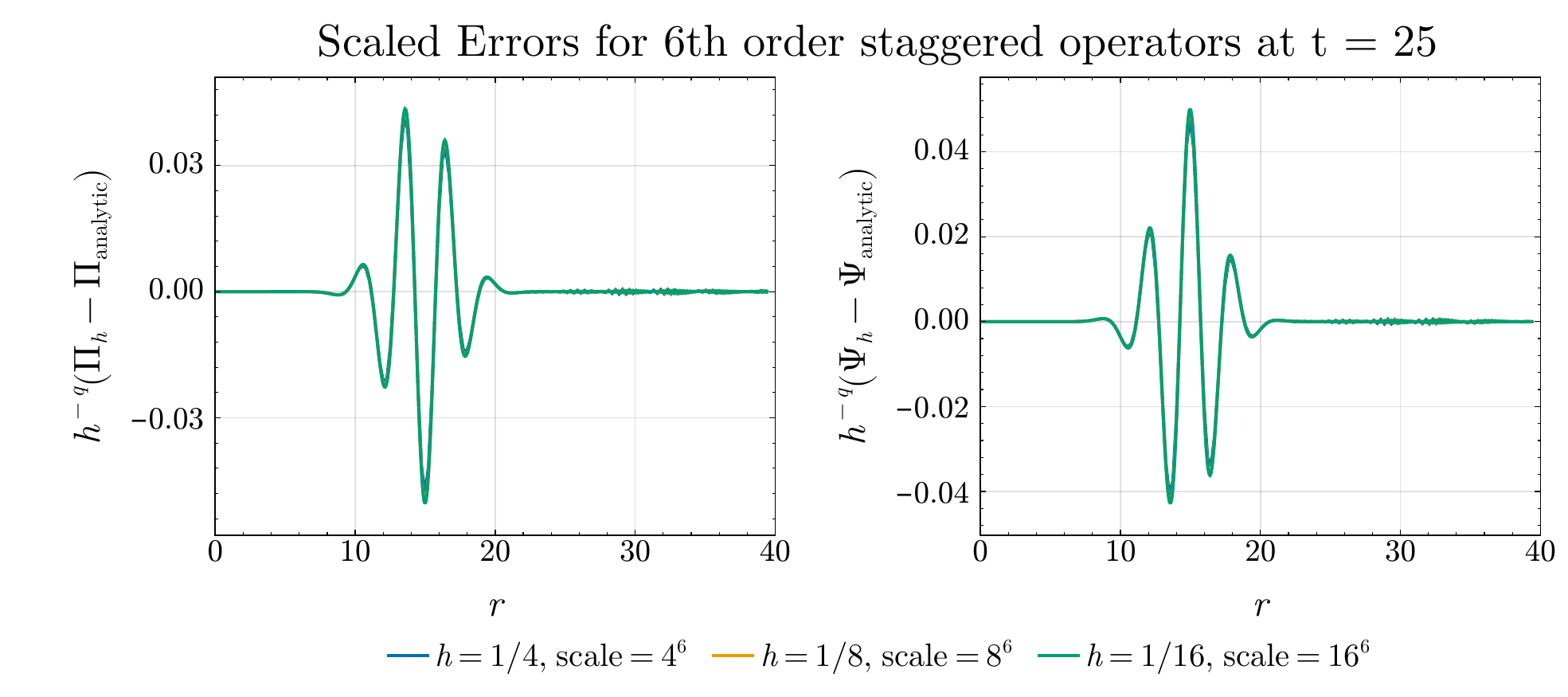}
    \caption{Convergence diagnostics for the 4th-order and 6th-order staggered SBP operator at $t=25$ for the resolutions $h=(1/4,1/8,1/16)$. Shows scaled errors (not the solution!), scaled by $h^{-4}$. This is the configuration $p=2$ on $r=[0,39.5]$. The noise amplitude for the highest resolution $h=1/16$ is at the level of $10^{-9}$ and converges away with increased resolution.}
    \label{fig:wave-convergence-staggered-late}
\end{figure}

\FloatBarrier

\section{Conclusion}

We describe a mechanism to construct summation-by-parts operators for finite differences in curvilinear coordinate system with singularities. Our ansatz generalizes ideas presented in \cite{Gundlach2013}. Starting from existing Cartesian SBP operators, we distinguish between a scalar and a vector-valued norm, allow the vector-valued norm to have non-diagonal elements, and choose a set of accuracy conditions near the origin that determines the elements of these operators. We then derive the covariant divergence operator from the SBP property. We show examples for sets of operators in three-dimensional spherical symmetry, demonstrating that these operators leads to excellent (accurate, smooth, higher-order convergent) solutions at and near the origin, even though we place a grid point directly onto the origin. We also briefly reviewed how to construct stable SBP operators for staggered grids and compare to these operators.

Our ansatz is applicable to operators for any tensor rank and any spacetime dimension.

We make the operators available at \texttt{SphericalSBPOperators.jl} repository \cite{vretinaris2026}.

\begin{acknowledgments}
    We thank Luis Lehner for interesting and useful discussions.
    SV is grateful for the hospitality of Perimeter Institute where part of this work was carried out.
    Research at Perimeter Institute is supported in part by the Government of Canada through the Department of Innovation, Science and Economic Development and by the Province of Ontario through the Ministry of Colleges, Universities, Research Excellence and Security.
    We acknowledge the support of the Natural Sciences and Engineering Research Council of Canada (NSERC).
    Nous remercions le Conseil de recherches en sciences naturelles et en génie du Canada (CRSNG) de son soutien.

    Plots were generated using \texttt{Makie.jl} \cite{DanischKrumbiegel2021}, time evolutions were performed with \texttt{OrdinaryDiffEq.jl} \cite{Rackauckas2017}. We used extended precision types from \texttt{MultiFloats.jl} \cite{MultiFloats.jl} and eigenvalues computed with \texttt{GenericSchur.jl} \cite{GenericSchur.jl}. The cartesian SBP operators were computed with \texttt{SummationByPartsOperators.jl} \cite{ranocha2021}.
\end{acknowledgments}

\bibliography{main.bib}

\end{document}